\documentclass[12pt]{article}

\usepackage{a4,graphicx,amsmath,amssymb}
\usepackage[verbose]{wrapfig}
\def\Bbb{\mathbb}
\def\BZ{\Bbb Z} 
\def\BC{\Bbb C} 

\catcode`@=11 \@addtoreset{equation}{section} \catcode`@=12

\begin{document}
\begin{titlepage}
\renewcommand{\thefootnote}{\fnsymbol{footnote}}
\begin{flushright}
hep-th/0703020\\
IITM/PH/TH/2006/17  \\
May 2007(v3)
\end{flushright}
\vspace{1.0cm}
\begin{center}
\large{\bf Chiral primaries in the Leigh-Strassler 
deformed ${\cal N}=4$ SYM -- a perturbative study }
\end{center} 
\bigskip 
\begin{center}
Kallingalthodi Madhu\footnote{E-mail: \texttt{madhu@physics.iitm.ac.in}} and 
Suresh Govindarajan\footnote{E-mail: \texttt{suresh@physics.iitm.ac.in}} 
\\  Department of Physics \\
Indian Institute of Technology Madras 
\\ Chennai 600036 INDIA\\ 
\end{center}

\begin{abstract}
We look for chiral primaries in the general Leigh-Strassler deformed 
${\cal N}=4$ super Yang-Mills theory by systematically computing the 
planar one-loop anomalous dimension for single trace operators up to 
dimension six. The operators are organised into representations of the 
trihedral group, $\Delta(27)$, which is a symmetry of the Lagrangian. We 
find an interesting relationship between the $U(1)_R$-charge of chiral 
primaries and the representation of $\Delta(27)$ to which the operator 
belongs. Up to scaling dimension $\Delta_0=6$ (and conjecturally to all 
dimensions) the following holds: The planar one-loop anomalous dimension
vanishes only for operators that are in the singlet or three 
dimensional representations of $\Delta(27)$. 
For other operators, the vanishing of the one-loop anomalous dimension occurs 
only in a sub-locus in the space of couplings. 
\end{abstract}
\end{titlepage}
\renewcommand{\thefootnote}{\arabic{footnote}}
\setcounter{footnote}{0}

\section{Introduction}

The AdS-CFT correspondence due to Maldacena \cite{Maldacena:1997re} is a 
concrete realisation of 't Hooft's proposal relating Yang-Mills (at 
large-$N$) to string theory. The correspondence relating ${\cal N}=4$ 
supersymmetric Yang-Mills (SYM) theory to type IIB strings propagating on 
$AdS_5\times S^5$ has been tested in many different ways. The first tests 
involved matching computations in the supergravity limit of the string 
theory and the corresponding ones in the conformal field theory (CFT)
following the proposal in 
\cite{Witten:1998qj}. Recently, it was realised that the anomalous 
dimensions of operators in this theory are given by the energy spectrum of 
a spin-chain\cite{Minahan:2002ve,Beisert:2003tq}.

The ${\cal N}=4$ theory has a high degree of symmetry and thus it is of 
interest to understand versions with lesser symmetry.  Orbifolds of this 
theory provide one obvious class of CFT's\cite{KS,LNV}. These lead to 
theories which are dual to type IIB string theory propagating on 
$AdS_5\times X^5$, where $X^5$ are five manifolds that are orbifolds of 
$S^5$. More generally, $X^5$ must be a Sasaki-Einstein manifold. A new 
class of such manifolds that lead to CFT's with ${\cal N}=1$ supersymmetry 
have also been constructed recently. These manifolds have been called 
$L_{pqr}$ spaces -- cones over these spaces are Ricci-flat non-compact six 
dimensional manifolds with $SU(3)$ holonomy and are natural 
generalisations of the conifold \cite{Martelli:2004wu,Cvetic:2005ft}.  
Thus, these are examples where both sides of the AdS-CFT correspondence 
are understood even though we do not fully understand string theory on AdS 
spaces.

From a field theoretic perspective, Leigh and Strassler\cite{Leigh:1995ep} 
considered (multi-parameter) marginal deformations of ${\cal N}=4$ 
supersymmetric Yang-Mills theory that preserve ${\cal N}=1$ supersymmetry 
and are conformal. We shall refer to these theories are known as 
the Leigh-Strassler (LS) theories. While there have been attempts 
\cite{Kulaxizi:2006zc} to understand the anticipated dual string theories 
for these field theories, the precise correspondence is not known in all 
generality. An important development due to Lunin and Maldacena 
\cite{Lunin:2005jy} was the construction of the gravity duals for the 
$\beta$-deformed ${\cal N}=4$ SYM which is a sub-class in the generalised 
LS family of ${\cal N}=1$ theories. Rational values of $\beta 
= m/n$ in the deformation turn out to be related to $(\mathbb{Z}_n \times 
\mathbb{Z}_n)$ orbifolds of $S^5$ with discrete 
torsion\cite{Douglas:1998xa,Douglas:1999hq,Berenstein:2000hy,Berenstein:2000ux}.
 
The most general Leigh-Strassler ${\cal N}=1$ preserving deformations of 
${\cal N}=4$ SYM is given (in superfields) by the following 
superpotential:
\begin{eqnarray}
W= i h\ \textrm{Tr} \Big( e^{i \pi \beta} \Phi_1 \Phi_2 \Phi_3
 - e^{- i \pi \beta} \Phi_1 \Phi_3 \Phi_2 \Big) 
+ \frac{ih'}{3} \  \textrm{Tr} \Big( \Phi_1^3 + 
\Phi_2^3
+ \Phi_3^3 \Big) \ .
\end{eqnarray}
The ${\cal N}=4$ limit occurs when $h=g$ and $\beta=h'=0$. The
$\beta$-deformed theory is obtained when $\beta\neq0$ and $h'=0$. 
For generic values of the
deformations, the global symmetry of the deformed theory is the trihedral
group $\Delta(27)$ with its centre being a $\BZ_3$ sub-group of the $U(1)_R$
symmetry. This is in contrast to the situation when $X^5=L_{pqr}$, where one
has a $U(1)^3$ symmetry  which is intimately related to the toric nature
of these spaces.

The LS deformations are parametrised by the four couplings: $g$ (the
Yang-Mills coupling constant), $h$, $h'$ and 
$q\equiv e^{i\pi\beta}$. 
While these are all marginal at the classical level, they 
are not all marginal in the quantum theory. However, it has been 
argued by Leigh and Strassler that in a subspace of the four-dimensional
space of
couplings, the theory is conformal. In particular, the vanishing the four 
beta functions is related to the vanishing of the anomalous dimension of
the scalars fields. The exact expression for the subspace 
is not known. However, it is known to two loops and is given by
\begin{equation}
|h|^2 \Big( 1\ +\ \frac{1}{N^2} (q - \bar{q})^2 \Big)\ +\ |h'|^2 \frac{N^2 -4}{2 N^2} =\ g^2\ .
\end{equation}
In the large $N$ limit that we pursue in this paper, the above condition 
simplifies to
\begin{equation}
|h|^2 + \frac{|h'|^2}{2}  = g^2\ .
\end{equation}
We also choose to work with real $\beta$ which is possible at the one-loop 
level.  In principle, this can be done at higher loops as well since 
the  imaginary part of $\beta$ can be gotten  rid of by redefining $h$. 

Perturbative studies of the $\beta$-deformed theory have been carried out 
by several authors\cite{Niarchos:2002fc,freed,zan1,Rossi:2005mr,zan2,zan3}. 
Chiral primaries of 
these theories are known well at least in the planar (large-$N$) 
case\cite{freed}.  
The ultraviolet finiteness of the $\beta$-deformed theory has been studied 
in ref. \cite{Khoze:2005nd} and a proof of its ultraviolet finiteness at the 
planar level has also appeared recently\cite{Rossi:2006mu,Ananth:2006ac}. 

For the general LS deformations, the anomalous dimensions for 
chiral 
operators with low values of scaling dimension has been carried out (using 
super-graphs) in ref. \cite{zan3} as well as ref. \cite{Bundzik:2005zg}. 
This paper focuses on a 
planar one-loop computation of anomalous dimensions of single trace 
operators in order to systematically search for operators that are 
protected in the LS theory.\footnote{The planar condition simplifies
things and
enables us to obtain concrete results for operators with dimensions up to
six. The complexity arises from the increase in the number of operators
that one has to consider. For instance, at $\Delta_0=6$ and $Q=0$, one has
to consider $46$ operators. It must be pointed out that for the 
$\beta$-deformed theory in the
planar limit, ref.
\cite{Beisert:2005if} has obtained an integrable
dilatation operator and diagonalized it using  a Bethe ansatz.}. 
In this paper, we systematically search for operators whose anomalous
dimensions vanish at planar one-loop in the LS theory. The operators
are organised into representations of $\Delta(27)$. We find that
protected operators appear in one of the three representations, ${\cal
L}_{0,0}$ or ${\cal V}_a$ depending on the value of the scaling
dimension, $\Delta_0$ modulo three.

The paper is organised as follows. In section two, we present some of the 
background needed for the paper. In particular, we present the F-term 
superpotential in component form and show that it contains double trace 
operators. In section three, we compute the anomalous dimensions for 
operators of the form $\textrm{Tr}(Z_1^kZ_2^lZ_3^m)$. We present the 
details of our computation as well as verify that all contributions that 
appear from non F-term interactions cancel and that our results are gauge 
independent. In 
section four, we compute the anomalous dimensions for operators up to
dimension six in the 
$\beta$-deformed theory as a preliminary to the computing in the general 
Leigh-Strassler theory. We verify that the results are consistent with 
expected results up to dimension six operators. In section five which
contains the main results of this paper, after 
organising the operators using the trihedral group, we compute the planar 
one-loop anomalous dimension in the Leigh-Strassler theory. We 
present our conclusions and outlook in section six.  Some technical 
details are relegated to the appendix for completeness.

\section{Background}

\subsection{The component Lagrangian for the LS theory}

We present here the details of the Lagrangian for the general
Leigh-Strassler deformation in component form. It turns out that unlike 
the ${\cal N}=4$ Lagrangian, this Lagrangian \textbf{cannot} be written in 
terms of a single trace though the superfield Lagrangian is written as a 
single trace\footnote{This result is implicitly present in the work of
Freedman and G\"ursoy\cite{freed}.}. 
There are some terms that can only be written as a double 
trace -- these terms are however suppressed by a power of $1/N$ but cannot 
be neglected in the large $N$ limit as we will see.

The simplest way to see the appearance of double trace operators is to 
consider trace identity (for $SU(N)$ generators in the fundamental 
representation):
\begin{equation}
\label{trform}
\textrm{Tr}(A T^{a}) \textrm{Tr}(B T^{a}) = \textrm{Tr}(A B) 
- \frac{1}{N} \textrm{Tr}(A) \textrm{Tr}(B)
\end{equation} 
Notice that both $F_i^a=\partial \bar{W}/\partial \bar{Z}_i^a$ and 
$\bar{F}_i^a=\partial W/\partial Z_i^a$ are both of the form 
$\textrm{Tr}(A T^{a})$ for some $A$. We thus see that $|F_i^a|^2$ cannot 
be written in single trace form (using the above identity) unless the 
operators $A$ and $B$ are traceless (as in the ${\cal N}=4$ limit). The 
F-term interactions (involving bosonic fields) for the 
LS theory is given by the potential\footnote{In this paper, we 
denote the bosonic component of the superfield $\Phi_i$ by $Z_i$ and
the $q$-deformed commutator is $[Z_1,Z_2]_q \equiv q Z_1 Z_2 - \bar{q} Z_2 Z_1$.}
\begin{eqnarray}
\label{beta1}
V_F(Z)= \textrm{Tr} \Big( |h'|^2 \bar{Z_1}^2 Z_1^2 +\ h \bar{h'} [Z_2,Z_3]_q \bar{Z_1}^2  
-  \bar{h} h' [\bar{Z_2},\bar{Z_3}]_q Z_1^2 
- |h|^2 [Z_2,Z_3]_q [\bar{Z_2},\bar{Z_3}]_q\Big)
\nonumber \\
 - \frac{1}{N} \Big[ |h'|^2\, \textrm{Tr}(\bar{Z_1}^2)\, \textrm{Tr} (Z_1^2 )
+\ h \bar{h'} \,\textrm{Tr}([Z_2,Z_3]_q)\, \textrm{Tr} (\bar{Z_1}^2)  - \bar{h} h' \,
\textrm{Tr}(Z_1^2)\, \textrm{Tr}([\bar{Z_2},\bar{Z_3}]_q)   \nonumber \\
-|h|^2\, \textrm{Tr}([Z_2,Z_3]_q)\, \big( \textrm{Tr} [\bar{Z_2},\bar{Z_3}]_q \big) \Big] 
 +\textrm{cyclic permutations}\hspace{1.65in}
\end{eqnarray}
The first line are the single trace operators while the last two lines are 
the double trace operators. When $h'=0$, one can see that the double trace 
terms are proportional to $(q- \bar{q})^2$ which vanishes when $q=\pm1$. 
Thus the ${\cal N}=4$ SYM theory does not have double trace terms in its 
component Lagrangian. Note that the double trace operators also do not exist
when the gauge group is $U(N)$. Since the D-terms are unaffected by the 
Leigh-Strassler deformations to the superpotential, they are the identical 
to the one obtained in the ${\cal N}=4$ theory (written in terms of ${\cal 
N}=1$ superfields). The detailed Lagrangian is given in appendix A.
 
\subsection{Symmetries of the LS theory}

The ${\cal N}=4$ SYM theory has a $R$-symmetry which is $SU(4)$. In the 
$\beta$-deformed theory, this is broken down to $U(1)^3$ -- each of the 
three scalars has charge one under only one of three $U(1)$'s with 
$U(1)_R$ being identified with the diagonal. In the 
Leigh-Strassler theory, the $U(1)^3$ is further broken down to 
$U(1)_R\times\BZ_3$. The LS theory 
has another symmetry given by the cyclic permutation ${\cal C}_3$ 
of the three scalar fields. This however, does not commute with the 
$U(1)_R\times \BZ_3$. The
trihedral group, $\Delta(27)\sim (\BZ_3\times 
\BZ_3)\rtimes{\cal C}_3$, is a discrete subgroup of $SU(3)\subset
SU(4)$ that captures the essential non-abelian nature. The centre of this
group is a $\BZ_3\subset U(1)_R$. The $U(1)_R$ charge which is
proportional to the scaling dimension for chiral primaries becomes a 
$\BZ_3$ valued charge. Specifically, in our conventions, the value of
the scaling dimension, $\Delta_0$, modulo three is the $\BZ_3$ charge.
Historically, 
the appearance of a 27 parameter non-abelian discrete subgroup was first 
noticed in ref.  \cite{Aharony:2002hx}.\footnote{We thank Ofer Aharony
for bringing this to our notice.}
Our attention to the appearance of the $\Delta(27)$ was drawn 
from ref. \cite{Wijnholt:2005mp} which attributed it to S. Benvenuti. 
We have found
the trihedral group  extremely useful in organising the chiral operators. 
For instance, it reduced the number of free parameters for an operator
at dimension six from $46$ to three different operators with 
$26$, $10$ and $10$ parameters.

One may ask whether $\Delta(27)$ remains a symmetry of the quantum
theory.\footnote{We thank Justin David for raising this question.} 
As argued by Leigh and Strassler, the full quantum effective potential
will continue to be of the same form as the classical one. In other
words, quantum corrections renormalised the coefficients $h$, $h'$ and
$\beta$. Thus, $\Delta(*27)$ will remain a symmetry of the
superpotential. For $\Delta(27)$ to be a symmetry in the quantum theory, 
however it is also
necessary that the K\"ahler potential also respects this symmetry. The
tree-level K\"ahler potential does respect the symmetry and one needs to
check if this remains true at higher orders in perturbation theory.
We have not resolved this issue completely and hope to report on this
in the future. Since our one-loop computation makes use of only
tree-level interactions, the use of $\Delta(27)$ is a valid one.
 
Another useful invariance of the action is the following:
\begin{equation}\label{permute}
\Phi_1 \leftrightarrow \Phi_2\ ,\ h \rightarrow -h\ , \ \beta
\rightarrow -\beta\ \textrm{and } h' \rightarrow h'\ .
\end{equation}
This is \textit{not} a symmetry since it acts on the couplings as well.
This however leads to restrictions on the possible renormalisation of
coupling constants.
 
\subsection{Propagators}

The propagators for the various fields are as follows,
where Feynman gauge has been chosen to write down the gauge propagator.
\begin{align}
\langle Z^a_i \bar{Z}^b_j\rangle \ = \ \delta_{ij} \delta^{ab}
\frac{1}{k^2} \ ,
&\qquad
\langle A^a_{\mu} A^b_{\nu}\rangle \ = \ - \delta^{ab} \frac{g_{\mu \nu
} }{2 k^2} \ ,
\nonumber \\
\langle \lambda^a \bar{\lambda}^b\rangle \ = \ - \delta^{ab} \ 
\frac{\sigma^{\mu} k_{\mu}}{2 k^2} \ ,
&\qquad
\langle \psi^a_i \bar{\psi}^b_j\rangle \ = \ -\delta^{ab} \delta_{ij} \
\frac{\sigma^{\mu} k_{\mu}}{k^2}\ .
\end{align}
We have explicitly verified the gauge independence of our results by working in the Landau
gauge as well.

\section{Anomalous dimension of $\mathrm{Tr}\big(Z_1^kZ_2^lZ_3^m)$}

In ${\cal N}=1$ gauge theories, it is known that holomorphy is the basis 
for certain
non-renormalisation theorems\cite{Grisaru:1979wc}. In order to prove properties that make 
use of holomorphy, one usually works in superfields and regularisation 
schemes that are compatible with holomorphy.\footnote{A much more modern
use of holomorphy and its relation to the Wilsonian effective action
is due to Seiberg\cite{Seiberg:1993vc}.} In our context where we are 
computing anomalous dimensions of operators involving scalars that arise 
from chiral 
superfields, holomorphy implies that the only interaction terms that 
contribute to the anomalous dimension are those that arise from $F$-terms 
as we will explicitly verify.

In computing the anomalous dimension of the operator ${\cal O}$, we 
compute the two-point function of this operator with its conjugate 
operator, which we denote by $\bar{\cal O}$ and study its singularity when 
the two operators are coincident. One expects
$$
\lim_{|x|\rightarrow0}\langle {\cal O}(x) \bar{\cal O}(0)\rangle 
\sim \frac{1}{|x|^{2\Delta_0}} - \frac{\gamma\log|x|^2}{|x|^{2\Delta_0}}\ ,
$$
where $\Delta_0$ is the naive scaling dimension of operator and $\gamma$ its 
anomalous dimension. Thus the anomalous dimension is computed extracting 
the logarithmic singularities and summing over all such contributions.

For the family of operators $\mathrm{Tr}\big(Z_1^kZ_2^lZ_3^m\big)$, we 
find that, at large $N$ (i.e., in the planar limit), the one-loop contribution 
to the 
anomalous dimension from all interactions take the following form (on 
using dimensional regularisation):
\begin{equation}
\frac{N^{k+l+m+1}}{256\pi^6|x|^{2(k+l+m)}} \Big(\frac1\epsilon +
3\log|x|^2+\textrm{constant}\Big)\times \textrm{a combinatoric factor}
\end{equation}
When the sum of all contributions is such that the coefficient of 
$\ln|x|^2$ vanishes, we obtain a candidate for the chiral 
primary. Recall that for chiral primaries, the scaling 
dimension is determined entirely by its $U(1)_R$-charge and hence 
should receive \textbf{no} corrections. For a true chiral primary,
$\gamma$ vanishes to all orders. So the vanishing of the planar one-loop
contribution to any operator does not imply that it is a chiral primary
since it could obtain contributions at higher orders. However, such
operators provide us with candidates for chiral primaries. 

\subsection{Cancellation of non F-term contributions}

Since we are working in component form, we need to explicitly verify that 
all non-holomorphic contributions to the anomalous dimensions of chiral 
fields cancel out.  These contributions should vanish irrespective of 
whether the operator is a chiral primary or not. While this is expected 
\cite{dhoker}, we use this computation as a non-trivial check of our 
results. Such contributions come from three kinds of terms:
\begin{itemize}
\item {\bf D-term:} Figures \ref{fig:1} and \ref{fig:2} arise from the D-term 
interaction vertex \\
$-(g^2/4) \sum_{i,j}\textrm{Tr}\big([Z_i,\bar{Z}_i][Z_j,\bar{Z}_j]\big)$.
\begin{figure}[h]
\centering
\includegraphics[height=1.3in]{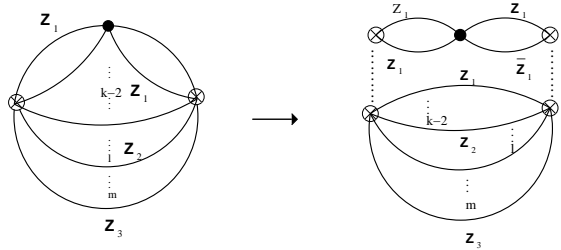}
\caption{Contribution from
$\textrm{Tr}\big([Z_1,\bar{Z}_1][Z_1,\bar{Z}_1]\big)$. The figure to the
right schematically shows how the logarithmic divergence was extracted.
The interaction vertex is labelled by a filled-in circle.}\label{fig:1}
\end{figure}
\begin{figure}[h]
\centering
\includegraphics[height=1in]{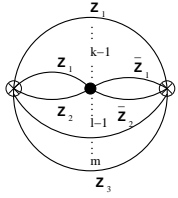}
\caption{Contribution from
$\textrm{Tr}\big([Z_1,\bar{Z}_1][Z_2,\bar{Z}_2]\big)$}\label{fig:2}
\end{figure}
\item {\bf Gluon exchange:} Figure \ref{fig:3} indicates the contribution from 
the
gluon-scalar interaction vertex $ig\textrm{Tr}\big(\partial_\mu
Z_i[A^\mu,\bar{Z}_i] + \partial_\mu \bar{Z}_i [A^\mu,Z_i]\big)$.
\begin{figure}[h]
\centering
\includegraphics[height=1in]{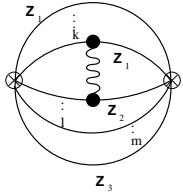}
\caption{Contribution from
gluon exchange}\label{fig:3}
\end{figure}
This diagram is gauge dependent and is logarithmically divergent in the 
Feynman gauge, but non-divergent in the Landau gauge. \item {\bf 
Self-energy:} Figure \ref{fig:4} indicates the contribution arising from 
the self-energy correction to all scalar propagators. This one is also a 
gauge dependent contribution.
\begin{figure}[h]
\centering
\includegraphics[height=1in]{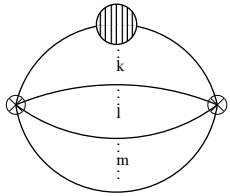}
\caption{Contribution from
corrections to the scalar propagator}\label{fig:4}
\end{figure}
\end{itemize}
As we will see the three contribution cancel for all operators that we are
considering here. We now provide some details of this cancellation. We have
also verified that the cancellation holds in both the Feynman and Landau
gauge though we will provide details for the Feynman gauge.

\subsection{Details of the cancellation}

\subsubsection*{Evaluation of Fig. \ref{fig:1}}

Fig. \ref{fig:1} has contributions coming from the interaction term $- 
\frac{g^2}{4} \textrm{Tr}\big( [Z_1,\bar{Z_1}][Z_1,\bar{Z_1}] \big)$. 
We evaluate the 
loop correction by doing one momentum integral and then taking the inverse 
Fourier transform to get the answer in position space.  We consider the 
fields pairwise and find the loop correction due to the interaction. We 
calculate loops involving interaction vertex separately in momentum space 
and then Fourier transform to position space. We multiply this with the 
contribution $1/|x|^{2(k+l+m-2)}$ from the part which does not involve 
interaction vertex. This is schematically explained in the diagram on the 
right in Fig. \ref{fig:1}

For an operator of general form $\textrm{Tr}\big(Z_1^k Z_2^l Z_3^m\big)$, 
there are $(k-1)$ contributions from this vertex. But when we have fields 
of only one flavor, say $Z_1$, there is an additional term giving a 
total of $k$ from this interaction vertex. However when $k =1$, there are 
no contributions from this vertex. Similar contributions for fields of 
other two flavors ensures that the combinatoric factor is symmetric in $k, 
l, m$. Hence the contribution from this diagram is
\begin{eqnarray}
\label{anom1}
  \frac{ 2 g^2 \cdot N^{k+l+m+1}\big(J_{klm}+G_{klm}\big) 
}{4 |x|^{2(k+l+m-2)}}\  \Big[ \int \frac{d^Dp}{(2\pi)^D} e^{i p\cdot x} \big( \int
\frac{d^Dk}{(2\pi)^D} \frac{1}{k^2 (k-p)^2} \big)^2 \ \Big] 
\end{eqnarray}
where $G_{klm}$ and $J_{klm}$ are the combinatorial factors with all the
properties described above, given by ($\delta_m$ is the Kronecker delta
function and is non-vanishing only when $m=0$)
\begin{eqnarray}
G_{klm} &=& -3 + \big(\delta_{k}+\delta_{l}+\delta_{m} \big) 
+ \big(\delta_{k}\delta_{l} + \delta_{l}\delta_{m} + 
\delta_{m}\delta_{k} \big)
- 3 \delta_{k} \delta_{l} \delta_{m} 
\end{eqnarray}
and
\begin{equation}
J_{klm} = k+l+m- \big(\delta_{k}\delta_{l} \delta_{m-1}
+\delta_{k-1} \delta_{l}\delta_{m} +\delta_{l-1} \delta_{m}\delta_{k} \big) 
\end{equation}
The momentum integral in Eqn.(\ref{anom1}) is evaluated in appendix.
We find the contribution from this figure as
\begin{eqnarray}
\label{d1}
&&\frac{g^2 N^{k+l+m+1} \big(J_{klm}+G_{klm} \big)}{2 (|x|^2)^{k+l+m-2}} 
\frac{ \Gamma^2[\epsilon] \Gamma^4[1 - \epsilon]}{(4 \pi)^D  
\Gamma^2[2 - 2 \epsilon]} 
\Big[ \int \frac{d^Dp}{(2\pi)^D} \frac{e^{i p\cdot x}}{(p^2)^{2\epsilon}}
 \Big] \nonumber \\
&=& \frac{g^2 N^{k+l+m+1}\big(J_{klm}+G_{klm} \big)}{256 \pi^6
|x|^{2(k+l+m)}} \Big( \frac{1}{\epsilon} + 1 + 5 \gamma_E + 3 \log\pi 
+ 3 \log|x|^2 \Big) \ ,
\end{eqnarray}
where $\gamma_E$ is Euler constant.

\subsubsection*{Evaluation of Fig. \ref{fig:2} } 

Fig. \ref{fig:2} involves the D-term interaction$ - \frac{g^2}{4}
\textrm{Tr}\big( [Z_1,\bar{Z_1}] [Z_2,\bar{Z_2}] \big)$ and
similar terms obtained by cyclic permutation of flavor indices. 
Here we notice that when the operator has fields of all three 
flavors the combinatorial factor must be 3. When the there are 
fields of two flavors this factor must be 2.
There is no such interaction when there are only fields of single 
flavor and hence there is no contribution from this diagram. The 
integral to be evaluated is the same as in Fig. \ref{fig:1}. The 
contribution from this interaction vertex is
\begin{equation}
\label{d2}
- g^2 \   \frac{N^{k+l+m+1}\ G_{klm}}{256 \pi^6 (|x|^2)^{k+l+m}} 
\Big( \frac{1}{\epsilon} +  1 + 5 \gamma_E + 3 \log\pi + 3 \log|x|^2 \Big) 
\end{equation}

\subsubsection*{Contribution from Fig. \ref{fig:3}} 

The interaction vertex is $ig\ \textrm{Tr} \partial_{\mu}Z [A^{\mu},
\bar{Z}]+\textrm{c.c}$. The calculation of corrections is again done by
taking propagators pairwise, as in the previous cases. We realise that out
of the different types of contractions possible only two are giving rise to
any divergence.  To find the combinatorial factor for Fig. \ref{fig:3} can
be easily identified as the some of the combinatorial factors of the above
two diagrams. The net contribution from this diagram is
\begin{eqnarray}
\label{d4}
&&2\ \frac{g^2}{2} \  \frac{N^{k+l+m+1}J_{klm}    
}{ (|x|^2)^{k+l+m-1}} 
\times \int \frac{d^Dp}{(2\pi)^D} e^{i p\cdot x} \int \int 
\frac{d^Dk d^Dq}{(2 \pi)^{2D}} 
\frac{k \cdot (k-p)}{(k-q)^2 (k-p)^2 q^2 (q-p)^2 k^2} \nonumber \\
&&= \frac{g^2 N^{k+l+m+1}J_{klm}}{256 \pi^6 (|x|^2)^{k+l+m}}  
\times \Big( \frac{1}{\epsilon} + 2 + 3 \gamma_E + 3 \log\pi + 3 \log|x|^2 \Big) 
\end{eqnarray}

\subsubsection*{Contribution from the self-energy}

This contribution arises out of the one-loop correction to the scalar
propagator $\langle Z_i \bar{Z}_i \rangle$. The calculation of one-loop
corrected scalar propagator is given in the appendix. This correction is
represented by the blob in Fig. \ref{fig:4}. Here again we calculate the
contribution from Fig. \ref{fig:4} by taking lines pairwise, where one of
the two lines has the blob. (This is needed to match the numerical factors
in the calculation of other diagrams.) This blob can appear on any of the
$\big(k + l + m \big)$ lines. Taking into consideration that the gauge group
is $SU(N)$ the combinatorial factor from Fig. \ref{fig:4} is seen to be
$J_{klm}$. Multiplying this by a factor of $\frac{1}{|x|^{2(k+l+m-2)}}$ from
the rest of the $(k + l + m - 2)$ lines, the one-loop correction to scalar
propagator is obtained in momentum space from Eqn. (\ref{betloop}) as
\begin{eqnarray}
&&- 2 N \big( |h|^2 +|h'|^2/2\big) \textrm{Tr}(T^a T^b) 
\int \frac{d^Dk}{(2\pi)^D}  \frac{1}{k^2 (k-p)^2 p^2} \nonumber \\
&=& - 2 N \big(|h|^2 +|h'|^2/2 \big) \textrm{Tr}(T^a T^b) 
\frac{\Gamma(\epsilon) B[1 - \epsilon, 1 - \epsilon]}{(4 \pi)^{2-\epsilon} 
(p^2)^{1+\epsilon}}
\end{eqnarray}
Inserting this into the Fig. \ref{fig:4} and calculating the loop
\begin{eqnarray}
\label{d31}
&&-2 \frac{ N^3 \big(|h|^2 +|h'|^2/2 \big) \Gamma(\epsilon) 
B[1 - \epsilon, 1 - \epsilon]}{(4 \pi)^{2-\epsilon}} 
\int \frac{d^Dp}{(2 \pi)^D}  \frac{1}{ (p-q)^2 (p^2)^{1+\epsilon}} \nonumber \\
&=& -2 \frac{ N^3 \big(|h|^2 +|h'|^2/2 \big) \Gamma(\epsilon) 
\Gamma(2 \epsilon) B[1 - 2 \epsilon, 1 - \epsilon] B[1 - \epsilon, 1 - \epsilon]}{(4 \pi)^{4-2 \epsilon}\ B[1 , 1 + \epsilon]\ \Gamma(2 + \epsilon)\  (q^2)^{2\epsilon}}
\end{eqnarray}
The $N^3$ factor arises when we contract the $\textrm{Tr}(T^a T^b)$ with
generators coming from the operator. In the above, the momentum integral is
again evaluated using Feynman parametrisation and then Fourier transformed
to position space to obtain
\begin{eqnarray}
-2 \frac{ N^{3} \big(|h|^2 +|h'|^2/2\big) \ 
J_{klm} }{256 \pi^6 |x|^4}
 \Big( \frac{1}{\epsilon} + 2 + 3 \gamma_E + 3 \log \pi + 3 \log |x|^2 \Big)
\end{eqnarray}
Together with the rest of the lines in Fig. \ref{fig:4}, 
which gives a factor of
$\frac{N^{k+l+m-2}}{|x|^{2(k+l+m-2)}}$ multiplying it,  
the total contribution of Fig. \ref{fig:4} is
\begin{equation}
\label{d3}
-2 \frac{ N^{k+l+m+1} \big(|h|^2 +|h'|^2/2\big)  J_{klm} 
}{256 \pi^6 |x|^{2(k+l+m)}}
 \Big( \frac{1}{\epsilon} + 2 + 3 \gamma_E + 3 \log \pi + 3 \log |x|^2 \Big)
\end{equation}
We can see that the coefficients of $\log|x|^2$ (as well as that of
$1/\epsilon$) in Eqns.(\ref{d1}), (\ref{d2}), (\ref{d4}), (\ref{d3}) add up
to zero. In particular, the term involving
$G_{klm}$ appears only from the contributions from Fig. 
\ref{fig:1} and \ref{fig:2} and they cancel. The term involving 
$J_{klm}$ adds up to give a term proportional to 
$( g^2 - |h|^2 - |h'|^2/2)$, which is proportional to the beta 
function and hence vanishes in the conformal limit. Hence the 
only contribution to the required correlator comes from the 
F-term interaction as expected.

\subsection{F-term contribution}

The computation of the anomalous dimension for quadratic operators such as
$\textrm{Tr}(Z_2Z_3)$ and $\textrm{Tr}(Z_1^2)$ differs from all
other values of $k,l,m$. Postponing the details for quadratic operators,
in the following subsection we will exclude  values of $k$, $l$, $m$ where
$k+l+m=2$.

The contribution from the F-term is obtained from a diagram similar 
to the one given in Fig. \ref{fig:2}. The interaction vertices 
involved are $|h|^2 \textrm{Tr}\big( [Z_1,Z_2]_q 
[\bar{Z_1},\bar{Z_2}]_q \big)$, $-|h'|^2 
\textrm{Tr}\big(\bar{Z_1}^2 Z_1^2 \big)$ and its cyclic permutations, 
where $[Z_1,Z_2]_q = q Z_1 Z_2 - \bar{q} Z_2 Z_1$. The 
combinatorics and integrals are the exactly as described earlier. 
In addition, here, when $k \neq 0,l=1,m=0$, 
contributions involving the parameter $q$ appear. The factor 
$S_{klm}$ is introduced to take this into account. The 
contribution to the above two-point correlator is
\begin{eqnarray}
 && \Big[ 2 |h|^2 \ \Big(\big( q^2 + \bar{q}^2 \big) 
 S_{klm} + G_{klm} \Big) 
 - \ 2 |h'|^2 (J_{klm}+G_{klm} ) 
\Big]\nonumber \\
   &\times& \frac{N^{k+l+m+1}}{256 \pi^6 |x|^{2(k+l+m)}} \Big( 
\frac{1}{\epsilon} +  1 + 5 \gamma_E + 3 \log( \pi) + 3 \log|x|^2 \Big) 
\end{eqnarray}
where
\begin{eqnarray}
S_{klm} &=& \delta_{k-1} \delta_{l} \big(1 - \delta_{m} \big) + \delta_{k} 
\delta_{l-1} \big(1 - \delta_{m} - \delta_{m-1} \big) \nonumber \\
&+& \delta_{l-1} \delta_{m} \big(1 - \delta_{k} \big) + \delta_{l} 
\delta_{m-1} \big(1 - \delta_{k} - \delta_{k-1} \big) \nonumber \\
&+& \delta_{m-1} \delta_{k} \big(1 - \delta_{l} \big) + \delta_{m} 
\delta_{k-1} \big(1 - \delta_{l} - \delta_{l-1} \big) 
\end{eqnarray}
The vanishing of the anomalous dimension now gives the condition
\begin{equation}
\label{anomcond}
\ |h|^2 \Big( \big( q^2 + \bar{q}^2 \big) 
S_{klm} + G_{klm} \Big) 
-\  |h'|^2 (J_{klm}+G_{klm} ) 
\Big) = \ 0 
\end{equation}

Before looking for solutions in full generality, let us first consider
the ${\cal N}=4$ limit, i.e., $q=\pm1$ and $h'=0$. In this limit, we
obtain
\begin{eqnarray}
\label{cond2}
2 S_{klm} + G_{klm} =\ 0
\end{eqnarray}
This has two non-trivial solutions:
\begin{enumerate}
\item[(i)] $k >2$, $l = m = 0$ and permutations thereof;
\item[(ii)] $k>1$, $l = 1$, $m = 0$ and permutations thereof.
\end{enumerate}
(i) corresponds to operators of the form $\textrm{Tr} \big( Z^k_1  \big)$,
 and (ii) corresponds to operators of the form $\textrm{Tr} \big( Z^k_1  Z_2
\big) $. All these are the known $\cal{N} =$ 4 chiral primary operators of
the form we considered\footnote{This list actually misses out the quadratic
operators $\textrm{Tr} \big(Z_iZ_j)$ which are also chiral primaries. As
mentioned earlier, the general formula given in Eqn. (\ref{anomcond}) is not
valid for these operators since there is an extra contribution appearing in
the deformed theory. This will be discussed in the next subsection.}.

We next consider the $\beta$-deformed theory which corresponds to 
keeping $h'=0$ and restoring arbitrary values for $q$.
The vanishing of the anomalous dimension is now
\begin{eqnarray}
\label{cond1}
\ |h|^2 \Big( \big( q^2 + \bar{q}^2 \big) 
S_{klm} + G_{klm} \Big) \ =\ 0
\end{eqnarray}
Among the two classes of solutions that we obtained in the ${\cal N}=4$
limit, we see that those of type (i) continue to have vanishing anomalous
dimension since $S_{klm}$ and $G_{klm}$ vanish separately for those values
of $k,l,m$. However, this is no longer true for the operators of type (ii).
These operators have the charges given in the list of chiral primaries given
by Lunin and Maldacena for the $\beta$-deformed theory\cite{Lunin:2005jy}.

Finally, we now consider the   Leigh-Strassler theory, where
$h'\neq 0$ as well. None of the ${\cal N}=4$ chiral primaries are protected
in this theory. However, in the limit $h =0$, the operator
$\textrm{Tr}\big(Z_1Z_2Z_3\big)$ (and also $\textrm{Tr}\big(Z_1Z_3Z_2\big)$) 
is found to be a solution of the Eqn. (\ref{anomcond}), which is easy to
understand as this operator cannot get any contribution at one loop
from the $h'$ interaction.
We will now discuss the 
operators $\textrm{Tr}\big(Z_1^2)$ and 
$\textrm{Tr}\big(Z_2Z_3\big)$ that were not considered earlier. 
We will see that both these operators are protected in the 
  Leigh-Strassler theory.

\subsection{Anomalous dimension for $\textrm{Tr}\big(Z_iZ_j)$}

The anomalous dimension for dimension two operators obtains contributions
from interactions involving double trace operators that appear in $V_F$.  In
the $\beta$-deformed theory, this interaction only affects the
$\textrm{Tr}(Z_2Z_3)$ operator, while in the general LS theory, the 
operator $\textrm{Tr}\big(Z_1^2)$ is affected by the $h'$ 
dependent double trace operator. For all other operators, one 
finds that interactions involving double trace operators provide 
contributions that are suppressed by a factor of $1/N$ relative 
to the single trace interactions and thus can be ignored in the 
large $N$ limit.

The computation for these operators differs from the above due to a subtlety
in taking the large $N$ limit. The deformed theories have an extra
interaction, as seen from Eqn. (\ref{beta1}), which is suppressed by a
factor of $N$ relative to other interactions as it is a multi-trace
operator\footnote{Recall that while the superfield Lagrangian has a single
trace, the component Lagrangian is obtained by eliminating the auxiliary
variables $D$ and $F$. Thus, the bosonic potential ends being a double trace
which can be rewritten as a single trace using identities such as Eqn.
(\ref{trform}). This term does {\bf not} appear for $U(N)$ as well. Note
also that it vanishes in the ${\cal N}=4$ limit.}.  For a dimension two
operator, the trace algebra works out as follows,
\begin{eqnarray}
&&\frac{1}{N}\, \mathrm{Tr}(T^{a} T^{b}) \mathrm{Tr}(T^{a} T^{b}) \mathrm{Tr}(T^{c} T^{d}) 
\mathrm{Tr}(T^{c} T^{d}) \nonumber \\
&&= \frac{1}{N}\, \Big[\frac{N^{2} - 1}{N} \times N \Big]^{2} = \frac{(N^{2}-1)^{2}}{N} \sim N^{3}
\end{eqnarray} 
The $\sim N^{3}$ contribution is seen to be of the same order as the one
from the single trace interaction piece in Eqn. (\ref{beta1}). This can be
ignored in the large $N$ limit while computing anomalous dimension for
operators of dimension $ > 2$.  The important point to note is that this
contribution is precisely the one that makes the anomalous dimension for
$\textrm{Tr}(Z_iZ_j)$ vanish (as has already been shown by others using
different methods.)

\subsection{Summary of results}

We have seen that in the $\beta$-deformed theory, at one-loop, the family of
operators of the form $\textrm{Tr}\big(Z_1^k\big)$ and
$\textrm{Tr}\big(Z_1Z_2\big)$ are protected. For the  
Leigh-Strassler theory, we have only two kinds of operators which survive on
including the $h'$ deformation. They are $\textrm{Tr}\big(Z_1^2\big)$ and
$\textrm{Tr}\big(Z_2Z_3\big)$. 

In order to further generalise the kinds of operators that one must
consider, we revisit the chiral primaries of the ${\cal N}=4$ theory. We
have found chiral primaries that involve only two flavors of the scalars.
What about those involving three flavours? The simplest one will involve one
of each flavor. The ${\cal N}=4$ chiral primary is
$$
\textrm{Tr} \Big( Z_1 Z_2 Z_3 + Z_1 Z_3 Z_2 \Big)\ .
$$
More generally, chiral primaries of ${\cal N}=4$ SYM are obtained by
considering linear combinations of all possible orderings of operators. For
instance, the above operator has $2=3!/3$ possibilities. The $3!$ is the
order of the permutation group in $3$ objects and the division by $3$
reflects the cyclic property of the trace. Given a monomial,
$z_1^{J_1} z_2^{J_2} z_3^{J_3}$, the corresponding  ${\cal N}=4$
primary is given by the expression (with $n=J_1+J_2+J_3$)
\begin{equation}
\sum_{\pi\in S_n} c_\pi\ \textrm{Tr}  \Big(\pi Z_1^{J_1} Z_2^{J_2}
Z_3^{J_3}\Big)\ ,
\end{equation}
where we sum over all permutations $\pi$ and $c_\pi$ is a symmetry
factor\cite{freed}. Thus, there is a one-to-one correspondence between
monomials and ${\cal N}=4$ chiral primaries. Thus, at dimension $\Delta_0$, the
number of chiral primaries is $(\Delta_0+1)(\Delta_0+2)/3$ which is the
number of monomials at degree $\Delta_0$.
Based on the form of F-term equations
like $\bar{F}_1 = q Z_2Z_3 - \bar{q} Z_3 Z_2=0$,
Freedman and G\"ursoy(FG) have argued that one needs to associate a
factor of $\bar{q}^2$ for terms that are related by the exchange of $Z_2$
and $Z_3$.
For instance, the chiral primary involving all three flavors,
will become
$$
\textrm{Tr} \Big( q Z_1 Z_2 Z_3 + \bar{q} Z_1 Z_3 Z_2 \Big)\ .
$$
in the $\beta$-deformed theory by their prescription. We will refer to
this as the FG prescription.  In the sequel, we will
verify the FG prescription works for operators involving up to
six powers of the scalars only when they turn out to be chiral
primaries.

\section{Chiral primaries in the $\beta$-deformed theory}

Chiral primaries in the $\beta$-deformed theory are classified by three
charges corresponding to a $U(1)^3$ subgroup of the $SO(6)$ R-symmetry in
the ${\cal N}=4$ theory. The three scalars $Z_1$,
$Z_2$ and $Z_3$ have charges $(1,0,0)$, $(0,1,0)$ and $(0,0,1)$ 
respectively. Chiral primaries are thus labelled by their 
$U(1)^3$ charges. Here we consider two-point functions of 
operators with charges $(J_1,J_2,J_3)$: $(2,1,1)$, $(3,1,1)$, 
$(2,2,1)$, $(4,1,1)$ and $(2,2,2)$ -- these are all the operators with 
$(J_1+J_2+J_3)\leq 6$ with all $J_i$ non-vanishing.

For a given choice of $(J_1,J_2,J_3)$, there are several operators that carry 
this charge. For example, there are three operators with charge 
$(2,1,1)$ as shown below. We choose linear combinations of all 
such operators in two steps. First, we obtain the corresponding 
${\cal N}=4$ primary. Second, we introduce powers of $\bar{q}$ 
following the FG prescription\cite{freed}. This is 
potentially a candidate for a chiral primary. We then put in 
arbitrary coefficients in front of all operators to make our 
ansatz more general. We then compute the anomalous dimensions of 
these operators at planar one-loop and obtain 
the condition for the vanishing of their anomalous dimensions.

In the following, we will see that the planar one-loop contribution to the
anomalous dimension for all operators can be written as the sum of the
absolute squares. This enables us to solve the equations easily without any
hidden assumptions.

\subsection*{$\Big(2,1,1\Big)$ operator}

We take the chiral primary to be of the following form
\begin{equation}
{\cal O}_{211} =  \textrm{Tr}\Big( Z_1^2 Z_2 Z_3 + b \bar{q}^2 Z_1 Z_2 Z_1 Z_3  
+ c \bar{q}^2 Z_2 Z_1^2 Z_3 \Big) \ .
\end{equation}
The computation of the anomalous dimension of 
this composite operator proceeds as before. The vanishing of the
anomalous dimension is given by the condition\footnote{It is easy to
extract the $3\times 3$ matrix of anomalous dimensions from the
following expression. It may have some use in writing out the
Hamiltonian for the spin-chain but its use is limited due to
the small length of the chain.}
\begin{equation}
\label{lbl1}
\Big\{ 
 3 |c|^2 + 4 |b|^2 + 3- 
2\, \mathrm{Re}\Big[ (2 \bar{b} + \bar{c}) 
+ 2 \bar{q}^2 b\bar{c}  \Big] 
\Big\}  =0
\end{equation}
The condition can be rewritten as follows:
$$
2|b-1|^2 + |c-1|^2 + 2|b \bar{q}^{2} -c|^2 \ = 0\ .
$$
This is a sum of three positive definite terms and is solved by $b=c=1$ and
$\bar{q}^{2}=1$ which makes it a chiral primary only for the ${\cal N} = 4$
theory.

\subsection*{$\Big(3,1,1\Big)$ operator}

\noindent
Here the chiral operator  is taken to be
\begin{equation}
{\cal O}_{311} =  \textrm{Tr}\Big( Z_1^3 Z_2 Z_3 + b \bar{q}^2 Z_1^3 Z_3 Z_2 
+ c \bar{q}^2 Z_1^2 Z_2 Z_1 Z_3  
+ d \bar{q}^4 Z_1^2 Z_3 Z_1 Z_2 \Big) \ .
\end{equation}
There is an ambiguity in applying the FG prescription. For instance, the
operator with coefficient $b$ can be associated with either $\bar{q}^2$
(as we have chosen) or $\bar{q}^6$. This ambiguity disappears when
$q^4=1$.
The condition for the vanishing of the planar one-loop anomalous dimension is
\begin{eqnarray}
\label{lbl2}
 3 |b|^2 + 4 |c|^2 
+ 4 |d|^2 + 3 - 2\, \mathrm{Re} \Big[  \bar{b} +  2 \bar{c} +  2 b \bar{d} q^4  
+ 2 c \bar{d}  \Big]  = 0\ .
\end{eqnarray}
The above expression  can be written as follows:
\begin{equation}
|b-1|^2 + 2|c-1|^2 + 2|b q^4 -d|^2 + 2|c-d|^2 =0\ .
\end{equation}
This has a solution only when $q^4 = 1$ and $b=c=d=1$. This is precisely
the situation where the FG prescription works. When $q=\pm1$, this
is ${\cal N}=4$ chiral primary. When $q=\pm i$, this operator is also a
protected operator. This is again an result expected from Lunin and
Maldacena\cite{Lunin:2005jy} -- this is a chiral primary in the
$\BZ_2\times\BZ_2$ orbifold of ${\cal N}=4$ theory. For all other values of
$\beta$, this operator is not a chiral primary.

\subsection*{$\Big(2,2,1\Big)$ operator}

For the operator
\begin{align}
{\cal O}_{221} =  \textrm{Tr}\Big( Z_1^2 Z_2^2 Z_3 
+ b \bar{q}^2 Z_1^2 Z_2 Z_3 Z_2 
&+ c \bar{q}^4 Z_1^2 Z_3 Z_2^2 
+ d \bar{q}^2 Z_1 Z_2 Z_3 Z_1 Z_2 \nonumber \\ 
&+ f \bar{q}^4 Z_2 Z_1 Z_3 Z_2 Z_1 + g \bar{q}^4 Z_1 Z_2^2 Z_1 Z_3
\Big) \ .
\end{align}
the  condition for the vanishing of the one-loop anomalous dimension is
\begin{eqnarray}
&& - 2\textrm{Re} \Big[  \bar{b} +   \bar{d} +  \bar{g} q^2  +  b \bar{c} 
+  b \bar{f} + 2 d \bar{f} +  d \bar{g} + \big(  b \bar{d}  
+  c \bar{f}  +  f \bar{g}  +  c \bar{g} \big) q^2   \Big] \nonumber \\ 
&&  + 4 |b|^2 + 3 |c|^2 + 5 |d|^2 
+5 |f|^2 + 4 |g|^2 + 3 = 0\ .
\end{eqnarray}
This can be written as the sum of squares as follows:
\begin{eqnarray}
&&|b-1|^2 + |d-1|^2 +|gq^2-1|^2 +
|b-c|^2 + |bq^2-d|^2 + |b-f|^2 \nonumber \\ 
&&+|cq^2-g|^2 + |cq^2-f|^2 + 2 |d-f|^2 +|d-g|^2 +|fq^2 -g|^2 =0 \ .
\end{eqnarray}
The  solution occurs only when $q^2 = 1$ and $b=c=d=f=g=1$
which is the known ${\cal N} =4$  chiral primary. Thus, this is not a chiral 
primary for generic values of $\beta$.

\subsection*{$\Big(4,1,1\Big)$ operator}

We next consider the operator with charge $(4,1,1)$. Below
the powers of $q$ have been assigned using the FG prescription.
However, there is an ambiguity in assigning the powers of $\bar{q}$. For 
instance, the operator multiplying the coefficient $b_3$ can be assigned
either $1$ or $\bar{q}^6$ since it can be reached by two different set of
exchanges. This ambiguity however goes away when $q^6=1$. 
\begin{align}
{\cal O}_{411} =  \textrm{Tr}\Big(b Z_1^4 Z_2 Z_3
+ b_1 \bar{q}^2 Z_1^4 Z_3 Z_2 &+ b_2 \bar{q}^4 Z_1^2 Z_2 Z_1^2 Z_3
 \nonumber \\
&+ b_3 \bar{q}^6 Z_1 Z_2 Z_1^3 Z_3 + b_4 \bar{q}^2 Z_1 Z_3 Z_1^3 Z_2
\Big) \ .
\end{align} 
The vanishing of the anomalous dimension at one-loop is
\begin{equation*}
\label{411}
 3 |b|^2 + 3 |b_1|^2  + 4 |b_2|^2 + 4 |b_3|^2 + 4 |b_4|^2 
-4 \mathrm{Re} \Big[  b \bar{b_4} +  \frac12 b \bar{b_1} 
+  b_1 \bar{b_3} q^6 +  b_2 \bar{b_3} +  b_2 \bar{b_4} \Big]
 = 0 \ .
\end{equation*}
Again, this can be written as the sum of absolute squares.
\begin{equation}
|b-b_1|^2 + 2 |b-b_4|^2 + 2 |b_1q^6-b_3|^2 + 2|b_2-b_3|^2 +
2|b_2-b_4|^2=0 \ .
\end{equation}
Clearly this has a solution $b=b_1=b_2=b_3=b_4$ only when $q^6=1$. Note that
this is precisely the value of $q$, where the ambiguity in the FG prescription
is removed.

\subsection*{$\Big(2,2,2\Big)$ operator}

For the operator
\begin{multline}
{\cal O}_{222} =  \textrm{Tr}\Big( d Z_1^2 Z_2^2 Z_3^2 
+ d_1 \bar{q}^2 Z_1^2 Z_2 Z_3 Z_2 Z_3 + d_2 \bar{q}^4 Z_1^2 Z_2 Z_3^2 Z_2 
+ d_3 \bar{q}^4 Z_1^2 Z_3 Z_2^2 Z_3  \\ 
+ d_4 \bar{q}^6 Z_1^2 Z_3 Z_2 Z_3 Z_2 + d_5 \bar{q}^8 Z_1^2 Z_3^2 Z_2^2 
+ d_6 \bar{q}^2 Z_3^2 Z_1 Z_2 Z_1 Z_2 
+ d_7 \bar{q}^4 Z_1 Z_2 Z_1 Z_3 Z_2 Z_3 \\
+ d_8 \bar{q}^6 Z_3^2 Z_2 Z_1 Z_2 Z_1 + d_9 \bar{q}^4 Z_1 Z_2^2 Z_1 Z_3^2 
+ d_{14} \bar{q}^6 Z_2^2 Z_1 Z_3 Z_1 Z_3
+ \frac{d_{11}}{2} \bar{q}^2 Z_1 Z_2 Z_3 Z_1 Z_2 Z_3 \\
+ d_{12} \bar{q}^4 Z_2 Z_1 Z_2 Z_3 Z_1 Z_3 + d_{13} \bar{q}^4 Z_1 Z_3 Z_1 Z_2 Z_3 Z_2 
+ d_{10} \bar{q}^2 Z_2^2 Z_3 Z_1 Z_3 Z_1
+ \frac{d_{15}}{2} \bar{q}^6 Z_1 Z_3 Z_2 Z_1 Z_3 Z_2 \Big) 
\end{multline}
we obtain the following condition for the vanishing of the anomalous dimension
\begin{equation}
\label{222}
\begin{split}
&3|d|^2+5 |d_1|^2 + 4 |d_2|^2  + 4 |d_3|^2 + 5 |d_4|^2 + 3 |d_5|^2 + 5 |d_6|^2 
+ 6 |d_7|^2 + 5 |d_8|^2 + 4 |d_9|^2 + 3 |d_{11}|^2 \\
& + 5 |d_{14}|^2 + 6 |d_{12}|^2  + 6 |d_{13}|^2 
+ 5 |d_{10}|^2 + 3 |d_{15}|^2  
-2\mathrm{Re} \Big[  d \bar{d_6} + d  \bar{d_1} +  d \bar{d_{10}} +  d_1 \bar{d_7} +  d_2 \bar{d_6} \\
& +  d_2 \bar{d_8} +  d_4 \bar{d_7} +  d_5 \bar{d_8} 
+  d_{14} \bar{d_{12}} +  d_6 \bar{d_9}  
+  d_8 \bar{d_9} +  d_1 \bar{d_3}  +  d_{13} \bar{d_{10}}  
+  d_{11} \bar{d_{13}}+  d_{12} \bar{d_{10}} +  d_{15} \bar{d_{13}} \\
&+  d_3 \bar{d_4} +  d_{14} \bar{d_{13}} 
+  d_5 \bar{d_4}   
+  d_6 \bar{d_7} +  d_1 \bar{d_2} +  d_4 \bar{d_2} +  d_{11} \bar{d_{12}} 
+  d_7 \bar{d_8}
+  d_{15} \bar{d_{12}} +  d_6 \bar{d_{12}}   \\
&+ d_9 \bar{d_{10}} +  d_9 \bar{d_{14}} +  d_8 \bar{d_{12}} +  d_4 \bar{d_{13}} + d_5 \bar{d_{14}} +  d_{10} \bar{d_3} +  d_3 \bar{d_{14}}
+  d_{11} \bar{d_7} +  d_{13} \bar{d_1} +  d_{15} \bar{d_7}  \Big]  =0 \ .
\end{split}
\end{equation}
This is independent of $q=e^{i\pi\beta}$ and is solved by $d = d_= \cdots = 1$
as can be clearly seen after rewriting the above expression in terms of 
sums of absolute squares.
\begin{equation}
\begin{split}
&|d-d_1|^2 + |d-d_6|^2 +|d-d_{10}|^2
+ |d_1-d_7|^2 + |d_1-d_3|^2 + |d_1-d_2|^2 +|d_1-d_{13}|^2 
\\
&
+|d_2-d_6|^2 + |d_2-d_8|^2 + |d_2-d_4|^2
+|d_3-d_4|^2 +|d_3-d_{14}|^2 + |d_3-d_{10}|^2
\\
&
+|d_5-d_8|^2 +|d_5-d_4|^2 +|d_5-d_{14}|^2+|d_4-d_7|^2 
+ |d_4-d_{13}|^2+|d_6-d_9|^2  
\\
&
+|d_6-d_7|^2+|d_6-d_{12}|^2 +|d_7-d_8|^2 + |d_7-d_{11}|^2 
+ |d_7-d_{15}|^2+|d_8-d_9|^2 
\\
&
+|d_8-d_{12}|^2 +|d_9-d_{14}|^2 + |d_9-d_{10}|^2 +|d_{11}-d_{13}|^2+|d_{11}-d_{12}|^2+|d_{14}-d_{12}|^2 
\\
&
+|d_{14}-d_{13}|^2
+|d_{12}-d_{10}|^2 +|d_{12}-d_{15}|^2
+|d_{13}-d_{10}|^2 + |d_{13}-d_{15}|^2=0 \ . \\
\end{split}
\end{equation}
Hence this is a chiral primary for any value of $\beta$ on
implementing the FG prescription.

By studying the non-renormalisation properties of operators up to dimension 
six, we see that for generic $\beta$, 
chiral primaries appear only as operators of the form $(k,k,k)$ and $(k,0,0)$
(other 
than the quadratic operators) as expected from the refs. 
\cite{freed,Berenstein:2000hy,Berenstein:2000ux,Lunin:2005jy}. Further, 
the FG prescription works for these operators. The absence of
an ambiguity in implementing the FG prescription seems to be the key
to the vanishing of the one-loop anomalous dimension. This also picks
out the special values of $q$ for which some operators are protected.

\section{General Leigh-Strassler deformation}

The general Leigh-Strassler deformation is invariant under the action of 
the trihedral group $\Delta(27)$ which is a finite non-abelian subgroup of 
$SL(3,\BC)$. The centre of this group is a $\BZ_3$ which is sub-group of 
$U(1)_R$. Thus, the $U(1)_R$ charge can be identified with the $\BZ_3$ 
charge.  Chiral primaries in this theory must appear as irreducible 
representations of $\Delta(27)$. In appendix B, we have provided relevant 
details of the irreducible representations of $\Delta(27)$. Based on the 
representation theory, we obtain the following important and useful 
result:
\begin{enumerate}
 \item When the scaling dimension, $\Delta_0=0$ mod $3$, then chiral
 primaries \textit{must} appear in any one of the nine one-dimensional
 representations, ${\cal L}_{Q,j}$ ($Q,j=0,1,2$). The representation ${\cal
 L}_{0,0}$ corresponds to a singlet of $\Delta(27)$. We will label such
 operators ${\cal O}_{\Delta_0}^{(Q,j)}$ to indicate the representation they
 belong to. The charge $Q$ for one-dimensional representations can be
 identified with the charge proposed in \cite{zan3}.
 \item When the scaling dimension, $\Delta_0=a$ mod $3$ ($a\neq0$), then
 chiral primaries appear in the three-dimensional representation, ${\cal
 V}_a$ and thus three operators form a triplet. We label all such operators
 by ${\cal O}_{\Delta_0}^a$. Given one operator of the triplet, the other
 two can be generated by the cyclic replacement $\tau:Z_1\rightarrow
 Z_2\rightarrow Z_3 \rightarrow Z_1$.
\end{enumerate}
This observation is useful in many ways. There will be no mixing between
operators which sit in distinct representations of $\Delta(27)$. This leads
to a nine-fold reduction in the operators that one needs to consider for 
one-dimensional
representations and a three-fold reduction for the three-dimensional
representations.

\subsection*{$\Delta_0 = 3$, $Q = 0$ operators}

Since $\Delta_0=0$ mod $3$, one has to only consider the one-dimensional
representations. There are three operators with $(Q,j)=(0,0)$ and we will
consider the most general linear combination of them.
\begin{equation}
{\cal O}_3^{(0,0)} =  \mbox{tr}\Big(a Z_1^3 + a Z_2^3 + a Z_3^3 + b Z_1 Z_2 Z_3 +
c Z_1 Z_3 Z_2  \Big) 
\end{equation}
The vanishing of the one-loop correction to the anomalous dimension is given
by 
\begin{eqnarray}
&&27|a|^2 |h'|^2 +  9  (h \bar{h'} q a \bar{b} + 
\bar{h} h' \bar{q} \bar{a} b) - 9  (h \bar{h'} \bar{q} a \bar{c} 
+ \bar{h} h' q \bar{a} c) \nonumber \\
&&- 3 ( |h|^2 \bar{q}^2 b \bar{c} + |h|^2 q^2 \bar{b} c)  + 3 (|b|^2 
+ |c|^2) |h|^2 \ =\ 0
\end{eqnarray}
This can easily be seen as equivalent to 
\begin{equation}
|\bar{h} \Big( b \bar{q} - c q \Big) + 3 a \bar{h'}|^2 \ =\ 0
\end{equation}
This has two solutions:
\begin{itemize}
\item[(i)] $a=0$, $b=q$ and $c=\bar{q}$. This implies that the $
\textrm{Tr}\big(q Z_1Z_2Z_3 + \bar{q} Z_1Z_3Z_2\big)$ which was a chiral
primary in the $\beta$-deformed theory is protected at one-loop in the
LS theory as well.
\item[(ii)] $a=1$, $b=-\frac{3\bar{h}'}{2\bar{h}} q$, $c=\frac{3\bar{h}'}{2\bar{h}} \bar{q}$.
This is the operator 
$$
\textrm{Tr}\big[(Z_1^3+Z_2^3+Z_3^3) - \frac{3m}2 
(q Z_1Z_2Z_3 -\bar{q} Z_1Z_3Z_2)\big]\ , \textrm{ where }\ m\equiv \frac{\bar{h}'}{\bar{h}}\ .
$$
\end{itemize}
There are two other operators with $Q=0$ and $j=1,2$ -- these are 
$$
{\cal O}_3^{(0,j)}=
\textrm{Tr}\big[Z_1^3+\omega^j Z_2^3+\omega^{2j} Z_3^3\big]\ .
$$ 
These are descendants\footnote{It follows from the representation theory 
of $\Delta(27)$ that there are nine descendants (with $\Delta_0=3$), one 
in each of the irreps, ${\cal L}_{Q,j}$. When, $(Q,j)=(0,0)$, there are 
three operators and one descendant while there are one operator in other 
sectors. We obtain two protected operators in the $\Delta_0=3$ sector 
which is consistent with this counting.} and hence are not chiral 
primaries.

\subsection*{$\Delta_0 = 3$, $Q = 1$ operators}

For this operator
\begin{equation}
{\cal O}_3^{(1,j)} =  \textrm{Tr}\Big( Z_1^2 Z_2 + \omega^j Z_2^2 Z_3 + \omega^{2j} Z_3^2 Z_1 \Big) 
\end{equation}
the vanishing of the one-loop correction to the anomalous dimension is
\begin{equation}
3\Big( |h'|^2 + |h|^2 |q-\bar{q}|^2 \Big) \ 
+\ 2\textrm{Re}\Big[ h \bar{h'} (q-\bar{q}) \Big(1+ \omega^j +
 \omega^{2j} \Big) \Big] = 0
\end{equation}
When $j\neq0$, the above equation has no solution (except in the ${\cal N}=4$ limit) 
implying that the operators are not chiral primaries. They are known descendants\cite{zan3}. However, when  $j=0$, the condition becomes
$$
3 |h' + h(q-\bar{q})|^2=0\ ,
$$
which has a solution only when $h'=h(q-\bar{q})$. At all other points in the space of couplings, the operator is a descendant.

\subsection*{$\Delta_0 = 4$ operator}

The operator that we will consider here is in the three-dimensional 
representation, ${\cal V}_1$, of $\Delta(27)$. Below, we will consider only 
one operator in the triplet since the result is valid for all three 
operators. The operator has charge $Q=1$. Further, it is a linear 
combination of several ${\cal N}=4$ primaries -- the unknown coefficients 
are labelled to remind the reader of this fact. For instance, below terms 
with coefficient using the same letter of the alphabet are part of the 
same ${\cal N}=4$ primary.
\begin{eqnarray}
{\cal O}_4^1 &=& \textrm{tr}\Big( Z_2^4 + b\ Z_1^3 Z_2 + c\ Z_1^2 Z_3^2 
+ c_1 Z_1 Z_3 Z_1 Z_3 + d\ Z_1 Z_2^2 Z_3 \nonumber \\
&&+\ d_1 Z_1 Z_2 Z_3 Z_2 + d_2 Z_3 Z_2^2 Z_1 + f Z_3^3 Z_2 \Big)
\end{eqnarray}
Requiring the vanishing of the one-loop correction to the anomalous dimension, we get,
\begin{eqnarray}
&&16 |h'|^2 + |b|^2 \Big( 2 |h'|^2 + |h|^2 |q-\bar{q}|^2 \Big) + 2 |c|^2 \Big( |h'|^2 + |h|^2 \Big)
+ |d|^2 \Big( |h'|^2 + 3 |h|^2 \Big) \nonumber \\
&&+\ |f|^2 \Big( 2 |h'|^2 + |h|^2 |q-\bar{q}|^2 \Big) + 4 |h|^2 |d_1|^2 +  8 |h|^2 |c_1|^2 
+ |d_2|^2 \Big( |h'|^2 + 3 |h|^2 \Big) \nonumber \\
&&+\ 2 \textrm{Re} \Big[ 4 h \bar{h'} q\ \bar{d} - 4 h \bar{h'} 
\bar{q}\ \bar{d_2} - h' \bar{h} (q-\bar{q}) b \bar{c}
+ h \bar{h'} q\ b \bar{d} - h' \bar{h} (q-\bar{q}) d_1 \bar{b} - 
h \bar{h'} \bar{q}\ b \bar{d_2} \nonumber \\
&&- h' \bar{h}  q\ c \bar{d} -\ 2 |h|^2 (c \bar{c_1}) (q^2 + 
\bar{q}^2) + h \bar{h'} q\ d_2 \bar{c} 
+ h \bar{h'}  (q-\bar{q}) c \bar{f} + 2 h' \bar{h} \bar{q}\ c_1 
\bar{d} \nonumber \\
&&- 2 h' \bar{h}  q\ c_1 \bar{d_2} +\ h' \bar{h} \bar{q}\ d 
\bar{f} - h' \bar{h} (q-\bar{q}) d_1 \bar{f} - h' \bar{h} q\ d_2 \bar{f}
- 2 |h|^2 q^2 \ d_1 \bar{d} \nonumber \\
&& - |h|^2 q^2 \ d_2 \bar{d} - 2 |h|^2 \bar{q}^2 \ d_1 \bar{d_2}  \Big] =\ 0
\end{eqnarray}
The above equation is rather hard to analyse and one may wonder if it has 
a solution. We now make use of the fact that in the limit $h'=0$, these 
equations should provide us conditions that appeared in the 
$\beta$-deformed theory. We have already seen that these can be written as 
the sum of squares. Using this result as input (and a check!), we deform 
the $h'=0$ term suitably such that all terms that appear as $h'\bar{h}$ 
that appear above are accounted for.  This strategy works rather well and 
we obtain an expression (given below) that is easily analysed.
\begin{eqnarray}
&& \ |4 \bar{h'} + \bar{h} \bar{q} d - \bar{h} q d_2|^2 + |\bar{h} \bar{q} d - \bar{h} q d_1 + \bar{h'} b|^2 
+ |\bar{h} \bar{q} d - \bar{h} q d_1 + \bar{h'} f|^2 \nonumber \\
&&+\, |\bar{h} \bar{q} d_1 - \bar{h} q d_2 + \bar{h'} b|^2 + |\bar{h} \bar{q} d_1 - \bar{h} q d_2 + \bar{h'} f|^2
+ |\bar{h} \bar{q} c - 2 \bar{h} q c_1 - \bar{h'} d|^2 \nonumber \\
&&+\, |\bar{h} q c - 2 \bar{h} \bar{q} c_1 + \bar{h'} d_2|^2
+ |\bar{h} (q-\bar{q}) b - \bar{h'} c|^2 + |\bar{h} (q-\bar{q}) f - \bar{h'} c|^2   = 0 \nonumber \\
\end{eqnarray}

This equation has the following definite solution providing us the 
required chiral primary operator at one-loop planar level.
\begin{eqnarray}
&&b = f = \frac{ 4 m^3 \,  (q^2 + \bar{q}^2 - 1)}{m^3  (q^2 + \bar{q}^2) 
- (q - \bar{q})^3 } \  \  \  \   
c = \frac{ 4 m^2 \, (q - \bar{q})\, (q^2 + \bar{q}^2 - 1)}{m^3 (q^2 + \bar{q}^2) - (q - \bar{q})^3} \nonumber \\
&&d = - \frac{ 4m \,  \big(m^3 \bar{q} - q (q - \bar{q})^2 \big)}{m^3  (q^2 + \bar{q}^2) - (q - \bar{q})^3} \  \  \  \   
c_1 = \frac{ 2 m^2 \,  \big(q - \bar{q} + m^3 \big)}{m^3 (q^2 + \bar{q}^2) - (q - \bar{q})^3} \nonumber \\
&&\  \  \  \  \  \  \  d_1 = \frac{ 4 m \,  (q - \bar{q}) \big(q - \bar{q} + m^3  \big)}{m^3 (q^2 + \bar{q}^2) - (q - \bar{q})^3}  \nonumber \\
&&\  \  \  \  \  \  \  d_2 = \frac{ 4 m \, \big(q^2 + \bar{q}^2 - 2 + m^3 q^3 \big)}{q^2 (m^3 (q^2 + \bar{q}^2) - (q - \bar{q})^3)}
\end{eqnarray}
where $m = \frac{\bar{h'}}{\bar{h}}$. We thus find only one protected operator
that exists for generic values of the couplings. 

\subsection*{$\Delta_0 = 6$, $Q = 1$ operators}

We consider dimension six operators which are in the one-dimensional 
representation of $\Delta(27)$. This is an example where the condition 
that the operator be in an irrep of $\Delta(27)$ (rather than an abelian 
subgroup as considered in \cite{zan1}, for instance) leads to a 
simplification. There is a three-fold reduction in the number of constants 
in the problem.
\begin{align}
{\cal O}_6^{(1,j)}  &=  \textrm{Tr}\Big( Z_1^5 Z_2 + b Z_1^4 Z_3^2 
+ b_1 Z_1^3 Z_3 Z_1 Z_3 + b_2 Z_1^2 Z_3 Z_1^2 Z_3 + c Z_1^3 Z_2^2 Z_3 
\nonumber \\
&+ c_1 Z_1^3 Z_2 Z_3 Z_2 + c_2 Z_1^3 Z_3 Z_2^2 + c_3 Z_1^2 Z_2 Z_1 Z_2 Z_3
+ c_4 Z_1^2 Z_2 Z_1 Z_3 Z_2 \nonumber \\
&+ c_5 Z_1^2 Z_2^2 Z_1 Z_3  
+ c_6 Z_1^2 Z_2 Z_3 Z_1 Z_2 + c_7 Z_1^2 Z_3 Z_1 Z_2^2 
+ c_8 Z_1^2 Z_3 Z_2 Z_1 Z_2 \nonumber \\
&+ c_9 Z_1 Z_2 Z_1 Z_2 Z_1 Z_3 \Big) \nonumber \\
&+\omega^j \textrm{Tr}\Big( Z_2^5 Z_3 
+ b Z_2^4 Z_1^2 + \ldots \Big) + \omega^{2j}\textrm{Tr}\Big( Z_3^5 Z_1 
+ b Z_3^4 Z_2^2 + \ldots \Big) 
\end{align}

Given the complexity of the expression for the anomalous dimension, 
we used the same strategy 
that was employed for the $\Delta_0=4$ operator. This enabled us to 
re-express the anomalous dimension as the sum of absolute
squares as given below.\footnote{The corresponding expression for the $j\neq0$
operators take a similar form. It is obtained by the the following 
replacements on the $j=0$ expressions: $m b\rightarrow
m b \omega^j$, $m b_i\rightarrow
m b_i \omega^j$, $m c\rightarrow m c \omega^{2j}$.
and $m c_i\rightarrow
m c_i \omega^{2j}$. Note that if any of the coefficients appears without
being multiplied by $m$, it is left unchanged. 
It turns out these equations do not have a
solution indicating that they are all descendants.}
\begin{align}
& \ | (q-\bar{q})  - m b|^2 + |m +  \bar{q} c_3 
-  q c_4|^2 
+ |m +  \bar{q} c -  q c_1|^2 \nonumber \\
&+ |m +  \bar{q} c_1 -  q c_2|^2 
+ |m +  \bar{q} c_6 -  q c_8|^2 
+ |m b +  \bar{q} c -  q c_5|^2 \nonumber \\
&+ |m c_7 +  \bar{q} c -  q c_3|^2 
+ |m c +  \bar{q} c_6 -  q c_1|^2
+ |m c_2 +  \bar{q} c_1 -  q c_4|^2 \nonumber \\
&+ |m c_5 +  \bar{q} c_8 -  q c_2|^2 
+ |m b +  \bar{q} c_7 -  q c_2|^2
+ |m b_1 +  \bar{q} c_3 -  q c_9|^2 \nonumber \\
&+ |m c +  \bar{q} c_3 -  q c_5|^2 
+ |m c_6 +  \bar{q} c_3 
-  q c_6|^2 + |m c_1 +  \bar{q} c_9 -  q c_4|^2 
 \\
&+ |m b +  \bar{q} c_6 -  q c_4|^2 
+ |m c_4 +  \bar{q} c_4 -  q c_8|^2 
+ |2 m b_2 +  \bar{q} c_5 -  q c_7|^2 \nonumber \\
&+ |m c_8 +  \bar{q} c_5 -  q c_9|^2 
+ |m c_1 +  \bar{q} c_6 -  q c_9|^2 
+ |m c_3 +  \bar{q} c_9 -  q c_7|^2\nonumber \\
&+ |m c_2 +  \bar{q} c_7 -  q c_8|^2  
+ |m b_1 +  \bar{q} c_9 -  q c_8|^2 
+ |m c +  \bar{q} b_1 -  q b|^2 \nonumber \\
&+ |m c_2 +  \bar{q} b -  q b_1|^2 
+ |m c_5 + 2  \bar{q} b_2 -  q b_1|^2 
+ |m c_7 +  \bar{q} b_1 - 2  q b_2|^2
\ =\ 0 \nonumber 
\end{align}
Trying to solve the constraint equations arising from the above, we can 
see that there are no generic state satisfying them. Thus, there are no 
protected operators for generic values of couplings. However, at specific 
sub-loci in the coupling space there are solutions. These belong to 
several branches which we list below: Branches (i) and (ii) are connected 
to the operators that appear when $q^4=1$ in the $\beta$-deformed theory. 
Branch (v) degenerates to a linear combination of ${\cal N}=4$ primaries 
when $q^2=1$. Branches (iii) and (iv) do not have such a limit.
\begin{itemize}
\item[(i)] $q^2=-1$, with $h$, $h'$ arbitrary. The solution is given by
$b=(q-\bar{q})/m$, $b_1=(m^2+2)/mq$, $b_2=q/m$; $c=c_2=c_4=c_5=c_6=c_7=1$,
$c_1=c_8=-(1+mq)$, $c_3=q(q+m)$ and $c_9=(mq-1-m^2)$. 
\item[(ii)] $q^2 = 1$, $mq=-1$. The solution is given by the
choices $b=c_1=c_8=0$,
$b_1=b_2= c = c_4 = c_5 = c_6=c_9=1$, $c_2=c_7=-1$ and $c_3=2$.
\item[(iii)] $m=1/q$. The solution is given by $b=c_1=c_3=q^2-1$, 
$b_1=c_2=c_4=c_6= c_7=c_9=1$, $b_2=1/q^2$, $c=q^4-q^2+1$, $c_5=(q^4-2)/q^2$
and $c_8=2/q^2$.
\item[(iv)] $m=-q$. The solution is given by $b=q^{-2}-1$, 
$b_1=c=c_4=c_5=ic_6=c_9=1$, 
$b_2=q^2$, $c_1=-1$, $c_2=q^{-4}-q^{-2}-1$, $c_3=2q^2$, 
$c_7=q^{-2}-2q^2$ and $c_8=-1+q^{-2}$.
\item[(v)] $m=q-\bar{q}$, $b=b_1=c=c_2=c_3=c_4=c_5=c_6=c_7=c_8=c_9=1$
and $b_2=1/2$.
\end{itemize}

\subsection*{$\Delta_0 = 6$,  $Q=0$ operators}

We now consider operators with $\Delta_0=6$ and $Q=0$. There are three 
operators in the representations, ${\cal L}_{0,j}$ that we need to consider.
We first consider the operator in the representation ${\cal L}_{0,0}$ -- it
consists of $46$ terms -- however, the number of independent coefficients is
reduced to $26$ due to the use of the trihedral symmetry.
We write the operator as the sum of four terms
\begin{equation}
{\cal O}_6^{(0,0)}={\cal O}_1 + \tau\big({\cal O}_1\big) + \tau^2\big({\cal O}_1\big)+ {\cal O}_2 \ ,
\end{equation}
where
\begin{align}
{\cal O}_1&= \textrm{Tr} \Big( a Z_1^6 + b Z_1^4 Z_2 Z_3 + b_1 Z_1^4 Z_3 Z_2 
+ b_2 Z_1^2 Z_2 Z_1^2 Z_3 + b_3 Z_1 Z_2 Z_1^3 Z_3  
\nonumber \\
& + b_4 Z_1^3 Z_2 Z_1 Z_3 + c Z_1^3  Z_2^3 + c_1 Z_1^2 Z_2 Z_1  Z_2^2 
+ c_2 Z_1^2  Z_2^2 Z_1 Z_2 + c_3 Z_1 Z_2 Z_1 Z_2 Z_1 Z_2 \Big)\ , \nonumber \\
{\cal O}_2&= \textrm{Tr}\Big(d\, Z_1^2 Z_2^2 Z_3^2 + d_1 Z_1^2 Z_2 Z_3 Z_2 Z_3 
+ d_2 Z_1^2 Z_2 Z_3^2 Z_2 + d_3 Z_1^2 Z_3 Z_2^2 Z_3 
\nonumber \\
&+ d_4 Z_1^2 Z_3 Z_2 Z_3 Z_2 
+ d_5 Z_1^2 Z_3^2 Z_2^2 + d_6 Z_3^2 Z_1 Z_2 Z_1 Z_2 
+ d_7 Z_1 Z_2 Z_1 Z_3 Z_2 Z_3
\nonumber \\
&+ d_8 Z_3^2 Z_2 Z_1 Z_2 Z_1 + d_9 Z_2^2 Z_1 Z_3^2 Z_1 
+ d_{10} Z_2^2 Z_3 Z_1 Z_3 Z_1 + d_{11} Z_1 Z_2 Z_3 Z_1 Z_2 Z_3  
\nonumber \\
&+ d_{12} Z_1 Z_2 Z_3 Z_1 Z_3 Z_2  + d_{13} Z_1 Z_2 Z_3 Z_2 Z_1 Z_3 + d_{14} Z_2^2 Z_1 Z_3 Z_1 Z_3 
+ d_{15} Z_1 Z_3 Z_2 Z_1 Z_3 Z_2\Big)\ , \nonumber
\end{align} 
and by $\tau({\cal O}_1)$ we mean the operator obtained by the cyclic 
replacement $\tau:Z_1\rightarrow Z_2\rightarrow Z_3 \rightarrow Z_1$ in all
the terms.  
The operators in the representations ${\cal L}_{0,j}$ ($j\neq0$) are 
given by
\begin{equation}
{\cal O}_6^{(0,j)}={\cal O}_1 + \omega^j \tau({\cal O}_1)+
\omega^{2j}\tau^2({\cal O}_1)\ .
\end{equation} 

After some long and rather tedious algebra, one  can express the vanishing
of the one-loop anomalous dimension for the operator ${\cal O}_6^{(0,0)}$
as the sum of absolute squares.
\begin{eqnarray}
&&3\, |6 a m - b_1  q + b  \bar{q}|^2 + |b_3 m - d_7  q + d_6  \bar{q}|^2 +
|b_4 m - d_8  q + d_7  \bar{q}|^2 \nonumber \\
&&+ |b_4 m - d_{4}  q + d_{13}  \bar{q}|^2 + |b_4 m - d_{4}  q + d_{7}  \bar{q}|^2 + |b_3 m - d_{7}  q + d_{1}  \bar{q}|^2 \nonumber \\
&&+ |b_1 m - 2 d_{15}  q + d_{7}  \bar{q}|^2 + |b m - d_{7}  q + 2 d_{11}  \bar{q}|^2 + |b_3 m - d_{13}  q + d_{10}  \bar{q}|^2 \nonumber \\
&&+ |b_4 m - d_{14}  q + d_{13}  \bar{q}|^2 + |b m - d_{13}  q + 2 d_{11}  \bar{q}|^2 + |b_1 m - 2 d_{15}  q + d_{13}  \bar{q}|^2 \nonumber \\
&&+ |b_3 m - d_{13}  q + d_{1}  \bar{q}|^2 + |b_1 m - 2 d_{15}  q + d_{12}  \bar{q}|^2 + |b m - d_{12}  q + 2 d_{11}  \bar{q}|^2 \nonumber \\
&&+ |b_3 m - d_{12}  q + d_{10}  \bar{q}|^2 + |b_4 m - d_{14}  q + d_{12}  \bar{q}|^2 + |b_4 m - d_{8}  q + d_{12}  \bar{q}|^2 \nonumber \\
&&+ |b_3 m - d_{12}  q + d_{6}  \bar{q}|^2 + |b m - d_{9}  q + d_{10}  \bar{q}|^2 + |b_1 m - d_{14}  q + d_{9}  \bar{q}|^2 \nonumber \\
&&+ |b_1 m - d_{8}  q + d_{9}  \bar{q}|^2 + |b_2 m - d_{5}  q + d_{8}  \bar{q}|^2 + |b_1 m - d_{8}  q + d_{2}  \bar{q}|^2  \\
&&+ |b_1 m - d_{14}  q + d_{3}  \bar{q}|^2 + |b m - d_{3}  q + d_{10}  \bar{q}|^2 + |b_2 m - d_{10}  q + d  \bar{q}|^2 \nonumber \\
&&+ |b_2 m - d_{1}  q + d  \bar{q}|^2 + |b m - d_{2}  q + d_{1}  
\bar{q}|^2 + |b_2 m - d_{6}  q + d  \bar{q}|^2 \nonumber \\
&&+ |b m - d_{3}  q + d_{1}  \bar{q}|^2 + |b m - d_{2}  q + d_{6}  \bar{q}|^2 + |b_1 m - d_{4}  q + d_{2}  \bar{q}|^2 \nonumber \\
&&+ |b_1 m - d_{4}  q + d_{3}  \bar{q}|^2 + |b_2 m - d_{5}  q + d_{4}  \bar{q}|^2 + |b m - d_{9}  q + d_{6}  \bar{q}|^2 \nonumber \\
&&+ |d_6 m - c_{1}  q + 3 c_{3}  \bar{q}|^2 + |d_{8} m - 3 c_{3}  q + c_{2}  \bar{q}|^2 + |d_{14} m - 3 c_{3}  q + c_{2}  \bar{q}|^2 \nonumber \\
&&+ 3 |d m - c  q + c_{2}  \bar{q}|^2 + 3 |d_{5} m - c_{1}  q + c  \bar{q}|^2 + 2 |d_{2} m - c_{2}  q + c_{1}  \bar{q}|^2 \nonumber \\
&&+ 2 |d_{9} m - c_{2}  q + c_{1}  \bar{q}|^2 + |d_{10} m - c_{1}  q + 3 c_{3}  \bar{q}|^2 + 2 |d_{3} m - c_{2}  q + c_{1}  \bar{q}|^2 \nonumber \\
&&+ |d_{4} m - 3 c_{3}  q + c_{2}  \bar{q}|^2 + |d_{1} m - c_{1}  q + 3 c_{3}  \bar{q}|^2 + 6 |c m - b_{4}  q + b  \bar{q}|^2 \nonumber \\
&&+ 6 |c_1 m - b_{2}  q + b_4  \bar{q}|^2 + 6 |c m - b_{1}  q + b_3  \bar{q}|^2 + 6 |c_2 m - b_{3}  q + b_2  \bar{q}|^2 \, = \, 0 \nonumber 
\end{eqnarray} 
The above equations lead to $52$ equations in $26$ unknowns. We find that
there are precisely \textit{two} solutions that exists for generic values of the
parameters. It is easy to see that there are
some identifications amongst the $d_i$. They are $d_1=d_6=d_{10}$, 
$d_2=d_3=d_9$, $d_4=d_8=d_{14}$ and $d_7=d_{12}=d_{13}$. This reduces the
number of unknowns to $18$. The explicit solutions are somewhat
unilluminating and will not be presented here. An important point
is that the two solutions can be characterised by $a=0$ and $a\neq0$.
In the limit, $h'\rightarrow 0$,
these two solutions reduce to the operators that exist in the
$\beta$-deformed theory for general values of $\beta$.
This is precisely
what happened for the $\Delta_0=3$, $Q=0$ operators as well. This is 
a clear indication that the operators that are protected at one-loop in
the $\beta$-deformed theory and are in the representation ${\cal L}_{0,0}$
survive the $h'$ deformation.
The operators in the representation ${\cal L}_{0,j}$ with $j\neq0$ 
are however not protected operators.

\subsection{Summary of results}

We have shown that it is useful to organise chiral primaries in terms of 
representations of the discrete group $\Delta(27)$. In particular, we
have seen that operators appearing only in one of
the three representations, ${\cal L}_{0,0}$ and ${\cal V}_a$ are
protected at planar one-loop level. We conjecture that this result is
true in general. The general pattern for operators protected at planar
one-loop (and possibly beyond) organised in terms of the trihedral 
group is given in the table below when $\Delta_0>2$. (We have excluded
the quadratic operators since they have a somewhat different behaviour.)
\begin{center}
\begin{tabular}{|l|c|c|c|}\hline
Scaling dim. & ${\cal N}=4$ theory & $\beta$-def. theory & LS theory \\ \hline
$\Delta_0=3r$ & ${\cal L}_{0,0}\oplus 
\frac{r(r+1)}{2} \big[\oplus_{i,j} {\cal L}_{i,j}\big]$ &
${\cal L}_{0,0}\oplus_{j} {\cal L}_{0,j}$ &
$2 {\cal L}_{0,0}$ \\[3pt] \hline
$\Delta_0=a$ mod $3$ & $\frac{(\Delta_0+1)(\Delta_0+2)}6$ ${\cal V}_a$ &
${\cal V}_a$ & ${\cal V}_a$ \\[3pt]\hline
\end{tabular}
\end{center}
The first column is only a reorganisation of the well-understood ${\cal
N}=4$ primaries into representations of $\Delta(27)$\cite{freed}. The
second column follows from the Lunin-Maldacena prediction that chiral
primaries in the $\beta$-deformed theory, for generic values of
$\beta$,
arise only with charges $(k,k,k)$ and $(k,0,0)$ rewritten in terms of
representations of $\Delta(27)$\cite{Lunin:2005jy}. The last column is based 
on our computations in the LS theory and has been verified up to and including
scaling dimension six. 

Further, we have seen that in other representations,
operators are only protected in a submanifold in the space of couplings.
These submanifolds consist of several branches, some of which do not
intersect the subspace of $\beta$-deformed theories.

\section{Conclusion}

In this paper, we have studied the planar one-loop contribution of
operators in the LS theory for operators up to dimension six. We have
used the trihedral group to classify the operators and this has lead to
a significant simplification to the problem. We find that for generic
values of couplings, the protected operators arise in the
one-dimensional representation, ${\cal L}_{0,0}$ when $\Delta_0=0$ mod
$3$ and in the three-dimensional representations ${\cal V}_a$ when
$\Delta_0=a$ mod $3$ ($a=1,2$). We conjecture that this is true in
general. It is interesting to see if there is a simple proof of this
statement.

The Leigh-Strassler superpotential makes an interesting appearance in a
different context. A recent computation all-orders perturbative
computation 
of the effective superpotential for
the so-called long-branes on the cubic torus using the topological
Landau-Ginzburg model turns out to be of the Leigh-Strassler
form\cite{Govindarajan:2006uy}. It is
possible that this computation may be related to the quantum effective
superpotential of the LS theory.\footnote{The authors of ref.
\cite{Dorey:2002pq} obtain an expression for the quantum effective
superpotential for the $\beta$-deformed theory using a relationship
with matrix models. It is also of interest to see if these two
effective potentials are related.} We are pursuing the
relationship of this work to the LS theory\cite{toappear}. In
particular, even if a direct map doesn't exist, it suggests a
re-ordering of the perturbative computation of the quantum effective
superpotential for the LS theory and that the renormalised coefficients
(up to an overall normalisation) should be expressible in terms of theta
functions of characteristic three. This statement is modulo the effect
of the the chiral Konishi anomaly which may modify the statement.

An open question is to find the gravity duals for LS theories. A more
limited question is to ask whether one can find special values of
the couplings like the case of rational $\beta$ in the $\beta$-deformed
theory. The crucial input in finding the gravity duals whenever $\beta$
was rational is the realisation that the effect of discrete torsion
in abelian orbifolds is to $q$-deform the ${\cal N}=4$ superpotential\cite{Douglas:1998xa,Douglas:1999hq,Berenstein:2000hy,Berenstein:2000ux}. 
One may ask whether discrete torsion in non-abelian orbifolds could
also produce the $h'$ deformation. The naive answer based on adapting 
the analysis of
ref. \cite{LNV} to include discrete torsion 
is that no such couplings can arise. However, since
those results are based on `dimensional reduction', it would be
interesting to actually carry out a CFT computation in string theory
to verify that such terms are not generated to come up with a no-go
theorem.

\subsection*{Acknowledgement:} One of us, KM, would like to thank Justin 
David for several helpful discussions throughout the work and HRI, Allahabad 
for 
hospitality during an extended  visit during the initial part of the work.  
We would also like to 
thank V. Ravindran and N. Dorey for useful discussions. The work of KM is 
supported by a Senior Research Fellowship from the CSIR (Award No. 
9/84(327)/2001-EMR-I). We also thank Justin David, S. Lakshmi Bala
and Prasanta Tripathy for a critical reading of a draft of this paper.

\appendix

\section{LS deformed ${\cal N} = 4$  Yang-Mills theory}

The Lagrangian density of the Leigh-Strassler theory in terms of
${\cal N}=1$ superfields is
\begin{eqnarray}
\label{zansup}
&{\cal L}& =\ \int d^2 \theta d^2 \bar{\theta} \textrm{Tr}\Big( e^{-g V} \bar{\Phi} 
e^{g V} \Phi \Big) 
+ \Big[ \frac{1}{2 g^2} \int d^2 \theta  \textrm{Tr} \Big( W^{\alpha} 
W_{\alpha} \Big) \nonumber \\
&+& i h \int d^2\theta  \textrm{Tr}  \Big( e^{i \pi \beta} 
\Phi_1 \Phi_2 \Phi_3
 - e^{- i \pi \beta} \Phi_1 \Phi_3 \Phi_2 \Big) 
+ \frac{ih'}{3} \int d^2 \theta  \textrm{Tr} \Big( \Phi_1^3 + \Phi_2^3
+ \Phi_3^3 \Big) + c.c. \Big] \nonumber \\
\end{eqnarray}
We denote the lowest component of the superfield $\Phi_i$ by $Z_i$ and its
fermionic partner by $\psi_{i}$. The vector multiplet in the Wess-Zumino
gauge has as components, the gauge field, $A_\mu$ and its superpartner,
the gaugino, $\lambda$ in addition to the auxiliary field $D$. All
fields transform in the adjoint of $SU(N)$.
Writing these in component fields we get the Lagrangian 
\begin{align}
\label{zan4}
{\cal L} &=  \textrm{Tr} \Big(\ -\frac{1}{2} F_{\mu \nu} F^{\mu \nu}  - 
i\ 2 \bar{\lambda} 
\sigma^{\mu} D_{\mu} \lambda  - i\ \bar{\psi_i} \sigma^{\mu} D_{\mu} \psi_i + 
D^{\mu}Z_i D_{\mu}\bar{Z_i} \nonumber \\
&- \frac{g^2}{4} [Z_i,\bar{Z_i}][Z_j,\bar{Z_j}]  + i \sqrt{2} g\ \bar{\psi_i}  
[Z_i,\bar{\lambda}] + i \sqrt{2} g\ \psi_i [\bar{Z_i},\lambda] \nonumber \\
&- i h\ \ \bar{\psi_3} [Z_1,\psi_2] +\ i \bar{h}\ \bar{\psi_3} [\bar{Z_1},\psi_2] 
- i h' Z_1 \psi_1 \psi_1 + i \bar{h'} \bar{Z}_1 \bar{\psi}_1 \bar{\psi}_1 + \textrm{cyc. perm.} \Big)\nonumber \\
&- V_F(Z)
\end{align}
where $D_{\mu}Z_i = \partial_{\mu}Z_i + i g [A_{\mu}, Z_i]$ and $V_F(Z)$ is as
given in Eqn. (\ref{beta1}). 

\subsection{Trace formulae for SU(N)}
Below, we provide the trace identities and normalisations that we have
used in our paper.
\begin{align}
&T^a T^a = \frac{N^2 - 1}{N} I \qquad  \qquad \textrm{Tr}(T^a T^b) 
= \delta^{ab} \nonumber \\
&\textrm{Tr}(A T^a B T^a) = \textrm{Tr}(A)  \textrm{Tr}(B) 
- \frac1N \textrm{Tr}(A B)  \\
&\textrm{Tr}(A T^a) \textrm{Tr}(B T^a) = \textrm{
Tr}(A B) 
- \frac1N \textrm{Tr}(A) \textrm{Tr}(B)  \nonumber
\end{align}

\section{Representations of the trihedral group, $\Delta(27)$}\label{trihedral}

In this appendix, we discuss the representation theory of the trihedral 
group $\Delta(27)$.\cite{RLeng,FFK,Bovier:1980gc} We expect chiral 
primaries of the Leigh-Strassler deformed theory to be in 
irreducible representations of this group. Trihedral groups are finite 
subgroups of $SL(3,\BC)$ of the form $A\rtimes {\cal C}_3$, where $A$ is a 
diagonal abelian group and ${\cal C}_3$ is the cyclic $\BZ_3$ generated by
$$
\tau =\begin{pmatrix} 0 & 0 & 1 \\ 1 &0  & 0 \\ 0 & 1 & 0 \end{pmatrix}\ .
$$
For the trihedral group, $\Delta(27)$, the group $A$ is generated by 
$g=\frac13(1,1,1)$ and $h=\frac13(0,1,-1)$\footnote{We denote by 
$\frac1R(a,b,c)$ the matrix 
$\textrm{Diag}(\epsilon^a,\epsilon^b,\epsilon^c)$ with $\epsilon$, a 
non-trivial $R$-th root of unity.}. In our example, $g$ turns out to be 
the centre of $\Delta(27)$ and is a subgroup of $U(1)_R$.

The irreducible representations of $\Delta(27)$ consist of nine 
one-dimensional representations, ${\cal L}_{Q,j}$ ($Q,j=0,1,2$) and two 
three-dimensional representations ${\cal V}_a$ ($a=1,2$). The charge under 
$g$ can be clearly identified with $U(1)_R$ charge.
\begin{enumerate}

 \item[$\mathbf{{\cal L}_{Q,j}}$] In the one-dimensional representations, one has
 the following action of the generators $h$ and $\tau$
 $$
 h \cdot v = \omega^Q\ v \ ,\quad  \tau \cdot v 
= \omega^j\ v \quad \textrm{ where } v\in 
 {\cal L}_{Q,j} \textrm{ and } \omega=e^{2\pi i/3}\ .
 $$
\item[$\mathbf{{\cal V}_a}$] In the three-dimensional representation, one has
$$
h\cdot \begin{pmatrix} v_0 \\ v_1 \\ v_2 \end{pmatrix}=
\begin{pmatrix} 1&0&0\\0& \omega^a & 0 \\ 0 & 0 & \omega^{2a} \end{pmatrix}
\begin{pmatrix} v_0 \\ v_1 \\ v_2 \end{pmatrix}
  \textrm{ and } \tau\cdot \begin{pmatrix} v_0 \\ v_1 \\ v_2 \end{pmatrix}
=\begin{pmatrix} 0 & 0 & 1 \\ 1 & 0 & 0 \\ 0 & 1 & 0 \end{pmatrix}
\begin{pmatrix} v_0 \\ v_1 \\ v_2 \end{pmatrix}
$$  
for $a=1,2$. Note that when $a=0$, the above representation is reducible 
to a direct sum, $\displaystyle{\oplus_{j=0}^2}{\cal L}_{1,j}$, of 
one-dimensional representations.
\end{enumerate}

The chiral superfields $(\Phi_1,\Phi_2,\Phi_3)$ are in the representation 
${\cal V}_1$ while their anti-chiral partners transform in the 
representation ${\cal V}_2$. We will choose our chiral primaries to be in 
one of these representations. Since the interactions in the 
Leigh-Strassler theory are invariant under $\Delta(27)$, it follows that 
chiral operators in distinct representations of $\Delta(27)$ cannot mix. 
However, operators in the same representation can and do mix.

\subsection{Polynomials as irreps of $\Delta(27)$}

On taking the commutative limit, all our chiral primary operators reduce 
to polynomials in three variables, $(z_1,z_2,z_3)$, where we replaced the 
matrices $Z_i$ by scalars $z_i$ (using the lower-case to indicate the 
replacement). First, the triplet $(z_1,z_2,z_3)^T$ transforms in the 
representation, ${\cal V}_1$. Second, we can organise the polynomials by 
degree -- the degree is the (naive) scaling dimension of the corresponding 
operator which we denote by $\Delta_0$. Thirdly, $\Delta_0$ mod $3$, is 
the $\BZ_3$ charge of the polynomial under the centre of the group (which 
is generated by $g$ defined above).

Polynomials in these variables of a given degree, $\Delta_0$, can be 
further organised into irreducible representations of $\Delta(27)$. The 
precise representation is decided by the value of $\Delta_0$ mod $3$. One 
has the following result:
\begin{itemize}

 \item \mbox{} [$\mathbf{\Delta_0=0}$ \textbf{mod} $\mathbf{3}$] All 
polynomials can be organised in one-dimensional representations of 
$\Delta(27$, i.e., ${\cal L}_{Q,j}$. For example, when $\Delta_0=3$ the 
polynomials $(z_1^3 + \omega^{j} z_2^3 + \omega^{2j} z_3^3)\in {\cal 
L}_{0,j}$ and $z_1z_2z_3\in {\cal L}_{0,0}$. In particular, there is no 
polynomial whose degree is $0$ mod $3$ that is in the representation 
${\cal V}_1$ or ${\cal V}_2$.
\item\mbox{} [$\mathbf{\Delta_0\neq0}$ \textbf{mod} $\mathbf{3}$] All 
polynomials must necessarily arise in one of the the 
three-dimensional representations. In fact, defining $a=\Delta_0$ mod $3$ 
with $a=1$ or $2$, the polynomials must be in the three-dimensional 
representation ${\cal V}_a$. For example, consider the F-term equations 
($dW=0$) which are of degree two. A straightforward analysis shows that 
the three equations are in the representation ${\cal V}_2$.
\end{itemize}
The proof of the above statements goes as follows. Note that the generator 
$g$ can be realised in terms of $h$ and $\tau$ as $h\tau^{-1} h^2 \tau$. 
Using this, notice that for a vector $v\in {\cal L}_{Q,j}$, $g\cdot v = 
\omega^{3Q} v = v$. Thus, the $\BZ_3$ charge associated with $g$ is zero 
implying that the $U(1)_R$ charge is zero modulo three. Similarly, for a 
triplet $\vec{v} \in {\cal V}_a$, $g \cdot \vec{v} = \omega^a\ \vec{v}$ 
implying that the $U(1)_R$ charge is $a$ modulo three.

This leads to the following conclusion: All chiral primaries with 
$\Delta_0=0$ mod $3$ must be in any one of the one-dimensional 
representations of $\Delta(27)$ while those where $\Delta_0=a$ mod $3$ 
must arise in the three dimensional representation ${\cal V}_a$.

\section{Integrals}

We evaluate dimensionally regulated momentum integrals, with $D = 4 - 
2\epsilon $, using the Feynman parametrisation
\begin{eqnarray}
\int \frac{d^Dk}{(2 \pi )^D} \frac{1}{(k^2)^{\alpha_1} 
(p^2)^{\alpha_2}}\  = \frac{1}{B[\alpha_1,\alpha_2]} \int_0^1 dx \int_0^1 dy\ 
\delta(1-x-y)\ x^{\alpha_1-1} y^{\alpha_2-1} \nonumber \\
 \times \int \frac{d^Dk}{(2 \pi )^D} \frac{1}{(x k^2 + y 
p^2)^{\alpha_1  + \alpha_2 }}\hspace{1.5in}  \\
=\ \frac{1}{B[\alpha_1,\alpha_2]} \int_0^1 dx\ x^{\alpha_1-1}  
(1-x)^{\alpha_2-1}\times \int \frac{d^Dk}{(2 \pi )^D}  
\frac{1}{(x k^2 + (1-x) p^2)^{\alpha_1 + \alpha_2 }} \nonumber
\end{eqnarray}
where $B[\alpha_1,\alpha_2]$ is the beta-function.

\noindent
The Fourier transform to position space is done using the following integral.
\begin{equation}
\label{frr}
\int \frac{d^Dp}{(2\pi)^D} \frac{e^{i p\cdot x}}{(p^2)^s} \ = \  
\frac{\Gamma[\frac{D}{2}-s]}{4^{\epsilon}\ 
\pi^{\frac{D}{2}}\Gamma[s]  (x^2)^{\frac{D}{2}-s}} \end{equation}

\noindent
{\bf Momentum integral (i)}
\begin{eqnarray}
&&\int \frac{d^Dk}{(2 \pi )^D} \frac{1}{k^2 (k-p)^2}\ =\ \int_0^1 dx \int 
\frac{d^Dk}{(2 \pi )^D} \frac{1}{\big( (k - px)^2 + p^2 x(1-x)\big)^2} 
\nonumber \\
&=&\int_0^1 dx \frac{\Gamma(\epsilon) x^{-\epsilon} 
(1-x)^{-\epsilon}}{(4 \pi)^{2-\epsilon} p^{2\epsilon}}\ 
=\ \frac{\Gamma(\epsilon) B[1-\epsilon,1-\epsilon]}{(4 \pi)^{2-\epsilon} 
p^{2\epsilon}}
\end{eqnarray}
Using the formula (\ref{frr}) above we obtain the integral in
Eqn.(\ref{anom1}) in position space as
\begin{eqnarray}
\int \frac{d^Dp}{(2\pi)^D} e^{i p\cdot x} \frac{1}{(p^2)^{2\epsilon}} = 
\frac{\Gamma[2 - 3\epsilon]}{4^{2\epsilon}\ \pi^{2-\epsilon}\  
\Gamma[2\epsilon]\ (|x|^2)^{2 - 3\epsilon}}
\end{eqnarray}
The integral in Eqn.(\ref{anom1}) gives
\begin{eqnarray}
\Big( \frac{\Gamma(\epsilon) B[1-\epsilon,1-\epsilon]}{(4 \pi)^{2-\epsilon} 
p^{2\epsilon}} \Big)^2 \frac{\Gamma[2 - 3\epsilon]}{4^{2\epsilon}\ 
\pi^{2-\epsilon}\ \Gamma[2\epsilon]\ (|x|^2)^{2 - 3\epsilon}} 
\end{eqnarray}
We expand this in powers of $\epsilon$ and obtain the answer in 
Eqn. (\ref{d1}).

\noindent
{\bf  Momentum integral (ii)}
The integral in Eqn. (\ref{d3}) using Feynman parametrisation
\begin{eqnarray}
&&\int \frac{d^Dp}{(2 \pi)^D}  \frac{1}{ (p-q)^2 (p^2)^{1+\epsilon}} = 
\frac{1}{B[1,1+\epsilon]} \int_0^1 dx \ x^{\epsilon}\int \frac{d^Dp}{(2 \pi)^D} 
\frac{1}{ \Big(x(p-q)^2 + xp^2 \Big)^{2+\epsilon}} \nonumber \\
&=& \int_0^1 dx\ x^{\epsilon} \Big(q^2 x (1-x)\Big)^{- 2 \epsilon} 
\frac{\Gamma[2 \epsilon]}{(4 \pi)^{2-\epsilon} \ B[1,1+\epsilon] 
\Gamma[2 + \epsilon]}\nonumber \\
&=& \frac{\Gamma[2 \epsilon] B[1-2\epsilon,  1-\epsilon]}{(4 
\pi)^{2-\epsilon} \ B[1,1+\epsilon] \Gamma[2 + \epsilon] (q^2)^{- 2 \epsilon}}
\end{eqnarray}
Again taking the Fourier transform of the above expression we get the
expression in Eqn. (\ref{d3}).

\noindent
{\bf Gluon exchange contribution to anomalous dimensions }

\begin{figure}
\centering
\includegraphics[width=3.5cm]{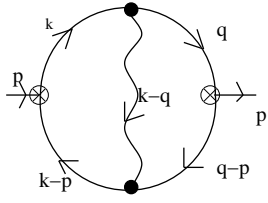}
\caption{Gluon exchange}\label{fig:5}
\end{figure}

\noindent
The contribution from interaction terms $ig\, \mathrm{tr}(\partial_\mu Z_i 
[A^\mu, \bar{Z}_i]_q)$ and $ig\, \mathrm{tr}(\partial_\mu \bar{Z}_i 
[A^\mu, Z_i]_q)$ to the anomalous dimension of the operator $\mathcal{O}$ 
is computed here. There are different contractions giving two distinct 
momentum integrals. The diagram in Fig. \ref{fig:5} can be evaluated when 
\underline{Case(1):} Both the interaction vertices involve 
$\partial_{\mu}Z$ (or $\partial_{\mu}\bar{Z}$ ), \underline{Case(2):} When 
one vertex involves $\partial_{\mu}Z$ and the other 
$\partial_{\mu}\bar{Z}$. We work out both the cases below.

\noindent
\underline{Case(1)}: 

From Fig. \ref{fig:5} given here we write down the Feynman integral for 
this case.
\begin{eqnarray}
&& \int \int \frac{d^Dk d^Dq}{(2 \pi)^{2D}} \frac{k \cdot (k-p)}{k^2 (k-q)^2 
(k-p)^2 q^2 (q-p)^2 } 
\end{eqnarray}
Considering only the part which is divergent we get,
\begin{eqnarray}
\int \int \frac{d^Dk d^Dq}{(2 \pi)^{2D}} \frac{1}{(k-q)^2 (k-p)^2
q^2 (q-p)^2 } \nonumber
\end{eqnarray}
We consider the $k$-integration first. Take $k' = k-p$. 
After dimensionally regulating and using Feynman parametrisation
\begin{eqnarray}
&& \int \frac{d^Dk'}{(2 \pi)^{D}} \frac{1}{{k'}^2 (k'-q + p)^2} = 
\int_0^1 dx \int \frac{d^Dk'}{(2 \pi)^{D}} \frac{1}{\Big((k'-(q-p)x)^2 
+ (q-p)x(1-x)\Big)^2}\nonumber \\
&=& \frac{\Gamma[\epsilon]}{(4 \pi)^{2-\epsilon} \Gamma[2] 
((q-p)^2)^{\epsilon}}\int_0^1 dx x^{-\epsilon} (1-x)^{-\epsilon} = \frac{\Gamma[\epsilon] B[1-\epsilon,1-\epsilon]}{(4 \pi)^{2-\epsilon} (q-p)^{2 \epsilon}}
\end{eqnarray}
Doing the $q$-integration in the same way,
\begin{eqnarray}
&&\frac{\Gamma[\epsilon] B[1-\epsilon,1-\epsilon]}{(4 \pi)^{2-\epsilon}}
\int \frac{d^Dq}{(2 \pi)^{D}} \frac{1}{{q}^2 \big((q - p)^2\big)^{1+\epsilon}} 
\nonumber \\
&=& \frac{\Gamma[\epsilon] B[1-\epsilon,1-\epsilon]}{(4 \pi)^{2-\epsilon}} 
\frac{1}{B[1,1+\epsilon]} \int_0^1 dy\ y^{\epsilon} \int 
\frac{d^Dq}{(2 \pi)^{D}} \frac{1}{\big((q - yp)^2 + p^2 y(1-y)
\big)^{2+\epsilon}} \nonumber \\
&=& \frac{\Gamma[\epsilon] B[1-\epsilon,1-\epsilon]}{(4 \pi)^{2-\epsilon}}
\frac{\Gamma[2 \epsilon]}{(4 \pi)^{2-\epsilon} B[1,1+\epsilon] 
\Gamma[2 + \epsilon] (p^2)^{2 \epsilon} } \int_0^1 dy\ y^{-\epsilon} 
(1-y)^{-2 \epsilon} \nonumber \\
&=& \frac{\Gamma[\epsilon] B[1-\epsilon,1-\epsilon]}{(4 \pi)^{4-2\epsilon}}
\ \frac{\Gamma[2 \epsilon] B[1-\epsilon,1-2 \epsilon]}{\Gamma[1+\epsilon] \ 
 (p^2)^{2 \epsilon} }
\end{eqnarray}
Again using formula (\ref{frr}) we Fourier transform and expand in powers of
$\epsilon$ to obtain
\begin{equation}
\frac{1}{256 \pi^6 |x|^4} \Big( \frac{1}{\epsilon} + 2 + 3 \gamma_E + 
3 \mbox{log}(\pi) + 3 \mbox{log}(|x|^2) \Big)
\end{equation}

\noindent
\underline{Case(2)} :
\begin{eqnarray}
&& \int \int \frac{d^Dk d^Dq}{(2 \pi)^{2D}} \frac{k \cdot (q-p)}{k^2 
(k-q)^2 (k-p)^2 q^2 (q-p)^2 } 
\end{eqnarray}
The divergent part of this is
\begin{eqnarray}
&& \int \int \frac{d^Dk d^Dq}{(2 \pi)^{2D}} \frac{k \cdot q}{k^2 
(k-q)^2 (k-p)^2
q^2 (q-p)^2 } \nonumber \\
&=& \frac{1}{2} \int \int \frac{d^Dk d^Dq}{(2 \pi)^{2D}} 
\frac{k^2 + q^2 - (k-q)^2}{k^2 (k-q)^2 (k-p)^2 q^2 (q-p)^2 } \nonumber \\
&=& \frac{1}{2} \int \int \frac{d^Dk d^Dq}{(2 \pi)^{2D}}  
\Big[ \frac{1}{(k-q)^2 (k-p)^2
q^2 (q-p)^2 } + \frac{1}{k^2 (k-q)^2 (k-p)^2 (q-p)^2} \nonumber \\
&& \ \ \ \ \ \ \ \ \ - \frac{1}{k^2 (k-q)^2 
q^2 (q-p)^2} \Big] \nonumber \\
&=&  \int \int \frac{d^Dk d^Dq}{(2 \pi)^{2D}}  \Big[ \frac{1}{(k-q)^2 (k-p)^2 q^2 (q-p)^2 } - \frac{1}{2 k^2 (k-q)^2 q^2 (q-p)^2} \Big]
\end{eqnarray}
The two integrals in the final expression above are already evaluated.
\begin{equation}
\Big( \frac{\Gamma[\epsilon] B[1-\epsilon,1-\epsilon] \Gamma[2 \epsilon] 
B[1-\epsilon,1-2 \epsilon]}{\Gamma[1+\epsilon] \ 
 } - \frac{\Gamma^4[1-\epsilon] \Gamma^2[\epsilon]}
{2\  \Gamma^2[2-2\epsilon] } \Big) \frac{1}{(4 \pi)^{4-2\epsilon} 
(p^2)^{2 \epsilon} }
\end{equation}

Fourier transforming to position space using formula (\ref{frr}) and 
expanding in powers of $\epsilon$, we get the value of the above integral 
to be $\frac{1}{256 \pi^6 |x|^2}$ Hence there is no contribution from this 
integral to the anomalous dimension.

\noindent
\section{One loop correction to scalar propagator }

\begin{figure}
\centering
\includegraphics[width=10.0cm]{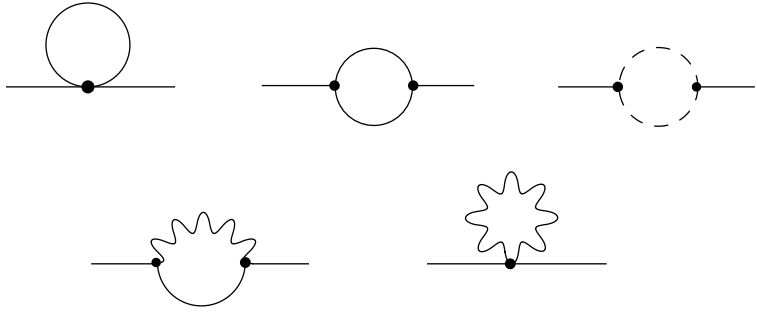}
\caption{One loop corrections to the scalar propagator}\label{fig:6}
\end{figure}
\vspace{0.5cm}

\noindent The diagrams in Fig. \ref{fig:6} are the one-loop contributions 
to the two point function $\langle Z_1 \bar{Z_1} \rangle$. The dashed 
lines are the fermionic propagators and the wiggly lines are gauge boson 
propagators. Gauge boson interactions are calculated in Landau gauge.

\noindent
{\bf Contribution from scalar tadpole} due to interaction term 
$-\frac{g^2}{4} \textrm{tr}( [Z_i,\bar{Z_i}][Z_j,\bar{Z_j}])$
\begin{eqnarray}
&& -\frac{g^2}{4} \cdot 2 \cdot \textrm{tr} \Big( ( T^b T^c -  T^c T^b ) \
 ( T^c T^a -  T^a T^c )\Big) \int d^Dk \ \frac{1}{p^2 \ k^2 \ p^2} \nonumber \\
&=& -g^2 \cdot \Big(  N tr(T^a T^b)\Big)
\int d^Dk \frac{1}{p^2 k^2 p^2} \nonumber \\
&=& -g^2 \ N tr(T^a T^b) \  \int d^Dk \frac{1}{p^2 k^2 p^2}
\end{eqnarray}
where $p_{\mu}$ is the external momentum and $D = 4 - 2\epsilon$.

\noindent
{\bf Contribution from scalar tadpole} due to interaction terms in $V_F(Z)$
\begin{eqnarray}
&&4 \Big[ |h|^2 \, \textrm{tr}\Big( (q T^b T^c - \bar{q} T^c T^b ) 
 (q T^a T^c - \bar{q} T^c T^a ) \Big) + |h'|^2 \textrm{tr}\big(T^b T^c T^d) 
\big(T^d T^c T^b) 
\Big] \nonumber \\
&& \  \  \  \  \  \  \times \int d^Dk  \frac{1}{p^2 \ k^2 \ p^2} \nonumber \\
&=&  -4 (|h|^2 + |h'|^2/2) \ N tr(T^a T^b) \  \int d^Dk \ \frac{1}{p^2 \ 
k^2 \ p^2}
\end{eqnarray}

\noindent {\bf Contribution from Yukawa interaction} vertices $i \sqrt{2} 
g\ \textrm{tr}\big( \psi_i [\bar{Z_i},\lambda] + \bar{\psi_i} 
[Z_i,\bar{\lambda}] \big)$.

The contribution from this interaction
\begin{eqnarray}
&&-(i \sqrt{2})^2 g^2 \ \textrm{tr}\big(T^c T^a T^d  - T^c T^d 
T^a \big)   \textrm{tr}\big( T^c T^b T^d - T^c T^d T^b \big) 
 \int d^Dk  \frac{\sigma^{\mu} k_{\mu} \   \sigma^{\nu} 
(p-k)_{\nu} }{2 p^2 k^2 (p-k)^2 p^2} \nonumber \\
&&= - 4 g^2 \ N tr(T^a T^b) \  \int d^Dk \  \frac{k\cdot p - 
k^2}{p^2 k^2 (p-k)^2  p^2}
\end{eqnarray}
Here we also use the identity $\textrm{tr} (\sigma^{\mu} k_{\mu} 
\sigma^{\nu} p_{\nu}) = 2 k\cdot p$

\noindent
{\bf  Contribution from Yukawa interaction} vertices 
$-ih \ \textrm{tr}\big( \psi_3  [Z_1,\psi_2]_q \big) - i h' Z_1 \psi_1 \psi_1 + \mathrm{c.c.} \big)$
\begin{eqnarray}
 \Big[ -(i )^2 |h|^2 \ \textrm{tr}\big(q T^c T^b T^d -  \bar{q} T^c T^d T^b \big)  
\textrm{Tr}\big( q T^c T^a T^d - \bar{q} T^c T^d T^a  \big) 
+ |h'|^2 \textrm{tr}\big(T^b T^c T^d) \big(T^d T^c T^b) 
\Big] \nonumber \\
\times \int d^Dk  \frac{\sigma^{\mu} k_{\mu} \  
\sigma^{\nu} (p-k)_{\nu} }{p^2 k^2 (p-k)^2 p^2} \hspace{3in} \\
= - 4 (|h|^2 + |h'|^2/2)\ N tr(T^a T^b) \  \int d^Dk \ 
\frac{k\cdot p - k^2}{p^2 k^2 (p-k)^2  p^2} \hspace{1in}\nonumber
\end{eqnarray}

\noindent
{\bf Contribution from scalar-gluon}  interaction terms $ i g\  
\textrm{tr}( \partial_{\mu}Z_i [A_{\mu}, \bar{Z}_i])$ ,  $i g\ 
\textrm{tr}( \partial_{\mu}\bar{Z}_i [A_{\mu}, Z_i])$
\begin{eqnarray}
&& 2 (i)^2 g^2 \ \textrm{tr} \big( T^b T^c T^d - T^b T^d T^c 
\big) \  \textrm{tr}\big( T^a T^c T^d - T^a T^d T^c \big) \nonumber \\
&& \  \  \  \  \  \  \  \  \  \ \int d^Dk  \Big[ \frac{ p^2 }{2 
p^2 k^2 (p-k)^2 p^2} \Big]\nonumber \\ 
&& + 2 (i)^2 g^2 \ \textrm{tr}\big( T^d T^c T^a - T^d T^a T^c 
\big) \  \textrm{tr}\big( T^d T^c T^b - T^d T^b T^c \big) \nonumber \\
&& \  \  \  \  \  \  \  \  \  \ \int d^Dk \Big[ \frac{ (p-k)^2}{2 
p^2 k^2 (p-k)^2 p^2}  \Big] \nonumber \\
&& + 4\ (i)^2 g^2 \textrm{tr}\big( T^d T^c T^a - T^d T^a T^c 
\big) \cdot \textrm{tr}\big( T^b T^c T^d - T^b T^d T^c \big) \nonumber \\
&& \  \  \  \  \  \  \  \  \  \ \int d^Dk \Big[ \frac{ p^2 - p 
\cdot k}{2 p^2 k^2 (p-k)^2 p^2} \Big] \nonumber \\
&&= \ 2 g^2 N \textrm{tr}(T^a T^b)  \int d^Dk \Big[ \frac{1}{2 p^2 \ k^2 \ p^2} + \frac{1}{p^2  k^2 (k-p)^2 p^2}\Big]
\end{eqnarray}

\noindent
{\bf Contribution from gluon tadpole} due to interaction term $- 
g^2 \textrm{tr}( [A_{\mu}, Z_i] [A_{\mu}, \bar{Z}_i])$ 
\begin{eqnarray}
&& - g^2 \cdot \textrm{tr}\Big( (T^c T^b - T^b T^c) \ (T^c T^a - 
T^a T^c) \Big)  
\int d^Dk \ \frac{ g_{\mu \nu}  
\, g^{\mu \nu}}{p^2 \ k^2 \ p^2} \nonumber \\ 
&& = -4 g^2 N \textrm{tr}(T^a T^b) \int d^Dk \ \frac{1}{p^2 \ k^2 \ p^2}
\end{eqnarray}

\noindent
Summing the contributions from each diagram, we see that quadratic divergences 
cancel. The one-loop correction to $\langle Z^a_i \bar{Z}^b_j \rangle$
is given as
\begin{eqnarray}
\label{betloop}
&& - (2 N) \ \textrm{tr} (T^a T^b) \cdot \big(|h|^2 + |h'|^2/2\big) 
\ \int d^Dk \ \frac{1}{ k^2 (p-k)^2 p^2} \nonumber \\
\end{eqnarray}
Fourier transforming we get
\begin{eqnarray}
&=&  - N \cdot \textrm{tr} (T^a T^b) \cdot \big(|h|^2 + |h'|^2/2 
\big) \ \big( Y_{122} + Y_{112} \big)
\end{eqnarray}
where $Y_{ijk}\ =\ \int d^4x \frac{1}{(x-x_i)^2 (x-x_j)^2 (x-x_k)^2}$.
In the large N limit we have $|h|^2 + |h'|^2/2 = g^2$. Hence we should get back 
the one-loop
correction in $\cal{N}$ = 4 theory. From Eqn. (\ref{betloop}) above, we 
get,
\begin{eqnarray}
2 N \cdot g^2 \ \textrm{tr} (T^a T^b) \cdot \big( Y_{122} + Y_{112} \big)
\end{eqnarray}
This expression is similar in struture
 with the one obtained in \cite{Klose:2003tw}.

\end{document}